# Optimizing hydrogen and e-methanol production through Power-to-X integration in biogas plants

Alberto Alamia[1]*, Behzad Partoon[2], Eoghan Rattigan[3] , Gorm Brunn Andresen[1]

[1]Aarhus University, Department of Mechanical and Production Engineering, Aarhus, Denmark
[2]Aarhus University, Department of Biological and Chemical Engineering, Aarhus, Denmark
[3]GreenLab Skive, Skive, Denmark

*Corresponding Author: alamia@mpe.au.dk

## Abstract

The European Union's strategy for achieving net zero emissions is heavily dependent on the development of hydrogen and e-/bio-fuel infrastructure and economy. These fuels are poised to play a critical role, functioning both as energy carriers and balancing agents for the inherent variability of renewable energy sources. Large-scale production will necessitate additional renewable capacity, and various Power-to-X (PtX) concepts are emerging in countries with significant renewable potential. However, sourcing renewable carbon presents a significant challenge in scaling the production of carbon-based e-fuels, and this is anticipated to become a limiting factor in the future. This investigation examines the concept of a PtX hub that sources renewable $CO_2$ from modern biogas plants, integrating renewable energy, hydrogen production, and methanol synthesis at a single site. This concept facilitates an internal, behind-the-meter market for energy and material flows, balanced by an interface with the external energy system. The size and operation of all plants comprising the PtX hub were co-optimized, considering various levels of integration with surrounding energy systems, including the potential establishment of a local hydrogen grid. The levelized cost of hydrogen and e-methanol were estiamted site commencing operation in 2030, taking in cosideration the recent legislation about renewable fuels of non-biological origin (RFNBOs). Our findings indicate that, in its optimal configuration, the PtX hub relies almost exclusively on on-site renewable energy, selling excess electricity to the grid for balancing purposes. The connection to a local hydrogen grid facilitates smoother PtX process operations, while the presence of a behind-the-meter market reduces energy prices, providing a buffer against external market variability. The results demonstrate the feasibility of achieving a levelized cost of methanol below 650 € /t and hydrogen production costs below 3 €/kg. In comparison, a standalone e-methanol plant would incur a 23% higher cost. The ratio of $CO_2$ recovered to methanol produced was identified as a critical technical parameter, with recovery rates exceeding 90% necessitating substantial investments in $CO_2$ and $H_2$ storage. Overall, our findings support the planning of PtX infrastructures that consider integration with the agricultural sector as a cost-effective pathway to access renewable carbon resources.

# 1 Introduction

The European Union's strategy for achieving a green transition has been significantly influenced by ambitious sustainability goals and the geopolitical instability caused by the Ukraine-Russia conflict. This situation has prompted the European Commission (EC) to substantially revise its green transition strategy, initially with the Fit for 55 package and subsequently with the more ambitious REPowerEU plan [1]. These initiatives aim to boost renewable energy while reducing the EU's reliance on imported hydrocarbons. The Fit for 55 package targets a 30% reduction in gas consumption by 2030, thereby lessening the need for hydrocarbon imports. Meanwhile, the REPowerEU plan sets more advanced targets for renewable energy and fuels, particularly emphasizing the production of hydrogen, e-fuels, and biomethane. Hydrogen is set to play a crucial role in the renewable-based energy system outlined by the REPowerEU plan and the recent legislation on Renewable fuels of non biological origin (RFNBOs) [1] . It will serve as an energy carrier for the industrial and transportation sectors and act as a balancing agent for the intermittent nature of renewable energy sources. The plan aims for the EU to produce 10 million tons of hydrogen annually by 2030 [1], addressing demands from the industrial sector and the production of electro-fuels for shipping, aviation, and fertilizers [1]. This ambitious target requires significant expansion of renewable energy and integration of Power-to-X (PtX) technologies, with an expected 65 GW of hydrogen production capacity from electrolysis units installed in the EU by 2030.

To align with the goal of moving away from fossil fuels, it is essential that the energy used in PtX processes is derived from renewable sources. This renewable energy can either come from surplus renewable energy, aiding in grid balancing [2], or from dedicated renewable sources adhering to the additionality principle [3]. Ensuring this alignment prevents the PtX demand from straining the existing electrical grid and allows PtX to play a pivotal role in balancing electricity demand and production, thus supporting the broader penetration of renewable energy and decarbonization efforts. Previous studies have highlighted various configurations for transmitting power, hydrogen, and natural gas to achieve a net-zero greenhouse gas emission EU system by 2050 [4]. Both the EC and major infrastructure operators advocate for establishing PtX sites and hydrogen transportation as key strategies for replacing imported natural gas. Public resistance to power grid expansion further supports the development of hydrogen grids [4, 5]. As hydrogen and e-fuels represent emerging markets driven by EU mandates, market conditions, including prices, regulations, and technical standards, are still developing. Nonetheless, numerous PtX projects are emerging in the EU, many with gigascale ambitions. In the absence of an existing hydrogen market with a transmission grid connecting producers and consumers, most projects focus on co-locating hydrogen production and conversion or planning in conjunction with upcoming hydrogen grids. Despite the energy conversion losses associated with e-fuel production, e-fuels are a primary driver for the developing PtX sector, particularly for decarbonizing significant markets like marine shipping and aviation [5]. Renewable methanol is versatile, serving as a shipping fuel, chemical feedstock, and intermediate chemical for kerosene synthesis. However, scaling up the production of carbon-based renewable e-fuels is challenged by the need for renewable carbon, which is expected to become a limiting factor in the future. Current processes converting biomass or waste to energy, such as biogas plants, biomass-fired plants, and waste-to-energy plants, provide the most readily available sources of biogenic carbon. Additional investments in $CO_2$ sequestration and gas cleaning are often necessary, except for biomethane plants, where such investments have already been made for biomethane production. Other sources of renewable carbon, including industrial fermentation, landfill gas, and direct air capture, are economically less viable for initial PtX development due to the higher cost of carbon sequestration [6]. The geographic location of biogenic carbon sources significantly affects the future establishment and cost-efficiency of e-fuel production facilities. Ideally, renewable carbon resources should be located near areas with high renewable energy (RE) potential. This strategic co-location eliminates the need for transporting hydrogen or electricity, thus reducing associated costs and mitigating grid balancing challenges [7].

One approach to address these particular challenges involves the integration of PtX hubs with biomethane facilities, typically found in rural areas abundant in renewable energy (RE) sources, as seen in regions like Denmark and northern Germany. The scale of e-fuel production in such PtX hubs is constrained to small-medium scale due to the availability of biogenic $CO_2$. However, these hubs are less likely to strain existing power grids and more likely to utilize excess power to stabilize local grid networks. While the presence of hydrogen infrastructure is not mandatory for e-fuel generation, its inclusion can enhance efficiency, making it an attractive addition. Another incentive for e-fuel synthesis at biomethane plants arises from their potential for $CO_2$ neutrality when utilizing $CO_2$ sourced from biogas upgrading processes, particularly from wet manure in closed digestate production, which aligns with Annex VI of the Renewable Energy Directive [8]. Despite the current limitations on scaling up biomethane production from manure in the EU [9], these $CO_2$ sources remain among the most sustainable and accessible options for e-fuel synthesis, regardless of regulatory frameworks, thus appealing to market participants seeking investment opportunities in e-fuels. The utilization of wet manure is widespread in Northern Europe, notably in Denmark [9] [10]. Given these advantages, integrating PtX technology with biomethane plants represents an initial step toward market consolidation, potentially supplemented by future large-scale e-fuel facilities utilizing $CO_2$ captured from biomass-fired and waste incineration plants. However, it's important to note that biogas reforming [11] is not currently considered a primary pathway for e-fuels or biofuels due to the expected demand for biomethane as a substitute for natural gas. Nonetheless, it could become a viable option if this demand diminishes in the future.

This investigation aims to provide an in-depth cost estimation of hydrogen and e-methanol produced in PtX hubs integrated with biomethane plants and to complement existing literature on cost estimation for renewable hydrogen and e-fuels [12] [13] [5] [14] [15]. This paper is grounded on the analysis of one of the first large-scale commercial projects of this kind, GreenLab Skive (GLS), located in northern Denmark. GLS functions as an industrial cluster emphasizing industrial symbiosis among various stakeholders. Products from one plant are utilized in another, minimizing waste and exemplifying an industrial symbiosis business models.

This study addresses several critical questions about the viability of PtX hubs:

1. Estimated production costs for hydrogen and e-methanol in the integrated PtX hub compared to standalone plants.
2. Identification of optimal topology, capacity, and operation of the PtX hub in relation to external energy prices, fossil $CO_2$ taxes, and the size of the biomethane plant.
3. Estimation of fair prices for behind-the-meter trading of energy and material flows within the PtX hub.
4. Estimation of the impact of add-on technologies to the PtX hub, such as district heating connection or pyrolysis with biochar carbon storage.

Unlike other techno-economic studies [16, 17] [18] [19] [20] [21] which focus mostly on process optimization, here we aim to simultaneously co-optimize the size of the plants in the PtX hub and their operation with hourly resolution. This optimization approach which is most used in transition studies of large energy systems [22] [23] allows to capture in the plant design and operation the combined impact of the variability in RE, legislation on RFNBOs and integration of the hub with the rest of the energy system.

The optimization model encompasses the investment for generation, conversion, storage, and the distribution of the main energy and material flows to achieve the least-cost outcome. As we aim to estimate the production costs in the initial large-scale rollout of the technology (2030 and afterwards) the optimization problem is driven by a mandate on demands yearly demand for $H_2$ and MeOH and not by market prices. The demand for methanol is varied to recover between 80% and 99% of the $CO_2$ separated from biogas. The demand for hydrogen defined two scenarios: large scale $H_2$ production (*$H_2$ to Grid*) with integrated MeOH production and *MeOH standalone* with $H_2$ only supporting the MeOH production. In the *H2 to Grid* scenario the hydrogen demand for the grid is set compatibly with the planned 100$MW_e$ GreenHyScale project in GLS (equivalent to 4000 full load hours). The availability of grid electricity for the PtX processes is set according to the latest RFNBOs legislation and all the cost assumption are forecasted for 2030 investment year. In the results, we compare the *$H_2$ to grid* and *MeOH standalone* scenarios, supplemented by a sensitivity analysis on key parameters. We also assess the impact of add-on technologies for this PtX configuration as connection to district heating grid and use of pyrolysis for biochar production and storage. It is worth noting that the capacity and operation of the biomethane plant are fixed in the model, and no cost is associated with ordinary operation of the process (Tab. 1).

**Table 1**: reference scenarios in the optimization of GLS PtX hub

| Scenario | $H_2$ production for Grid | Methanol production | $CO_2$ tax | Energy prices year | Biomethane | PtX configuration |
|---|---|---|---|---|---|---|
| *$H_2$ to Grid* | 272 (GWh/y) | 51-70 (GWh/y) | 0-150-250 (€/ton) | 2019 & 2022 | 190 (GWh/y) | Integrated |
| *MeOH standalone* | 0 (GWh/y) | 51-70 (GWh/y) | 0-150-250 (€/ton) | 2019 & 2022 | 190 (GWh/y) | Stand-alone |

# 2 Background

This section discusses the context of this study, providing a review of existing literature and of similar PtX hubs, followed by a discussion of the legal framework and finally description PtX hub concept at GreenLab Skive, as depicted in Figure 1.

## 2.1 Review previous literature

In recent years, the techno-economic assessment of PtX processes increased in relevance [24]. Most of the literature focuses on the impact of key technology parameters, as technolgoy of the electrolyzer [25] and their effieicny [26] [18] [19] and design of the MeOH syntehsys [16, 17] [27]. Typically these studies focus on the simulaiton of the PtX processes and partially on their optimization often assuming average electricity prices [20] [21] [28] [24]. With the intorduction of the RFNBOs legislation the impact of usign on-site renewables with higher variability and lower electricity cost has become more relevant. Studies including these aspects are [29] which focuses on German energy market conditions, and [30] which adresses similar conditons focusing on the forecatsing of electricty grid prices and not including the impact of RFNBOs legislaiton in the capacity optimization. Nevertheless, none of these studies co-optimized the capacity and operation fo the plants in the PtX site, econconpassing the benefits of the industiral symbiosys with behind-the-meter trading.
The estimated produciton cost of MeOH and $H_2$ varies among these studies, however it fits with the range indicated by the consotiums in industry sector [12] [31]. Presently, the production cost for renewable hydrogen is estimated between 3 and 10 €/kg [12] [31], a range primarily influenced by the capital costs of electrolyzers and the levelized cost of renewable electricity. If large-scale deployment materializes, the production cost for electrolyzers is expected to decrease, akin to the trends witnessed in wind power and solar PV in recent decades. Depending on location, renewable hydrogen could emerge as the most cost-effective method for hydrogen production by the end of the decade, with estimated costs potentially falling below 3 €/kg in the EU [12, 24]. As comparison the levelized cost of hydrogen production from fossil-based sources was approximately 1.5 €/kg for the EU [32], before Russia's invasion of Ukraine. The estimated production cost for e-methanol highly influenced by the cost for renewable hydrogen and assumption on the cost of carbon. The current estimate ranges between 800 and 1500 €/t [33]. For comparison fossil methanol, priced at 350-500 €/$t_{MeOH}$ [33] [24], would require a $CO_2$ tax of 180-230 €/$t_{CO2}$ on the estimated life cycle emissions (110 g/$MJ_{MeOH}$) [13] to reach a market price similar to e-methanol.

## 2.2 Overview of existing projects for e-MeOH

The flexibility of methanol as intermediate chemical and combustion fuel, makes it a key production for the sustainability transition. There are two main pathways for converting $CO_2$ to methanol using renewable electricity [34] [16] [35] [36]. One pathway is to reduce $CO_2$ to carbon monoxide (CO) and use convectional catalyst for methanol synthesis, in a two-step process. A second and more efficient pathway is direct hydrogenation of $CO_2$ with hydrogen over a heterogeneous catalyst through a one-step process that converts $CO_2$ directly to liquid fuels. CO2 hydrogenation is been studied and recently demonstrated at commercial scale by the Carbon Recycling Institute (CRI)'s George Olah plant and the Mitsui Chemicals in Japan and Mitsubishi Hitachi Power Systems Europe GmbH (MHPSE) [16] [26]. The EU commercial processing pioneering MeOH production from $CO_2$ is the George Olah Renewable Methanol Plant located in Iceland operated since 2012 by Carbon Recycling International. The plant produces 4kt/y of methanol, a size compatible with the $CO_2$ available from medium size biomethane plants.

**Table 2** - Overview of e-methanol plants in EU (commercial or engineering phase)

| Company | Location | Startup | Capacity (t/y) |
|---|---|---|---|
| CRI | Grindavik (IS) | 2012 | 4000 |
| HIF global | Punta Arenas (CL) | 2022 | 0.6 |
| GreenLab Skive | Skive (DK) | 2026 | 8000 |
| European Energy/Miotsui | Kasso (DK) | 2025 | 32000 |
| Ørsted | Ornskoldsvik (SE) | 2025 | 55000 |
| Total Energies | Leuna (DE) | 2025 | 1000 |
| ZATs | Zella-Mehelis (DE) | 2025 | 7000 |

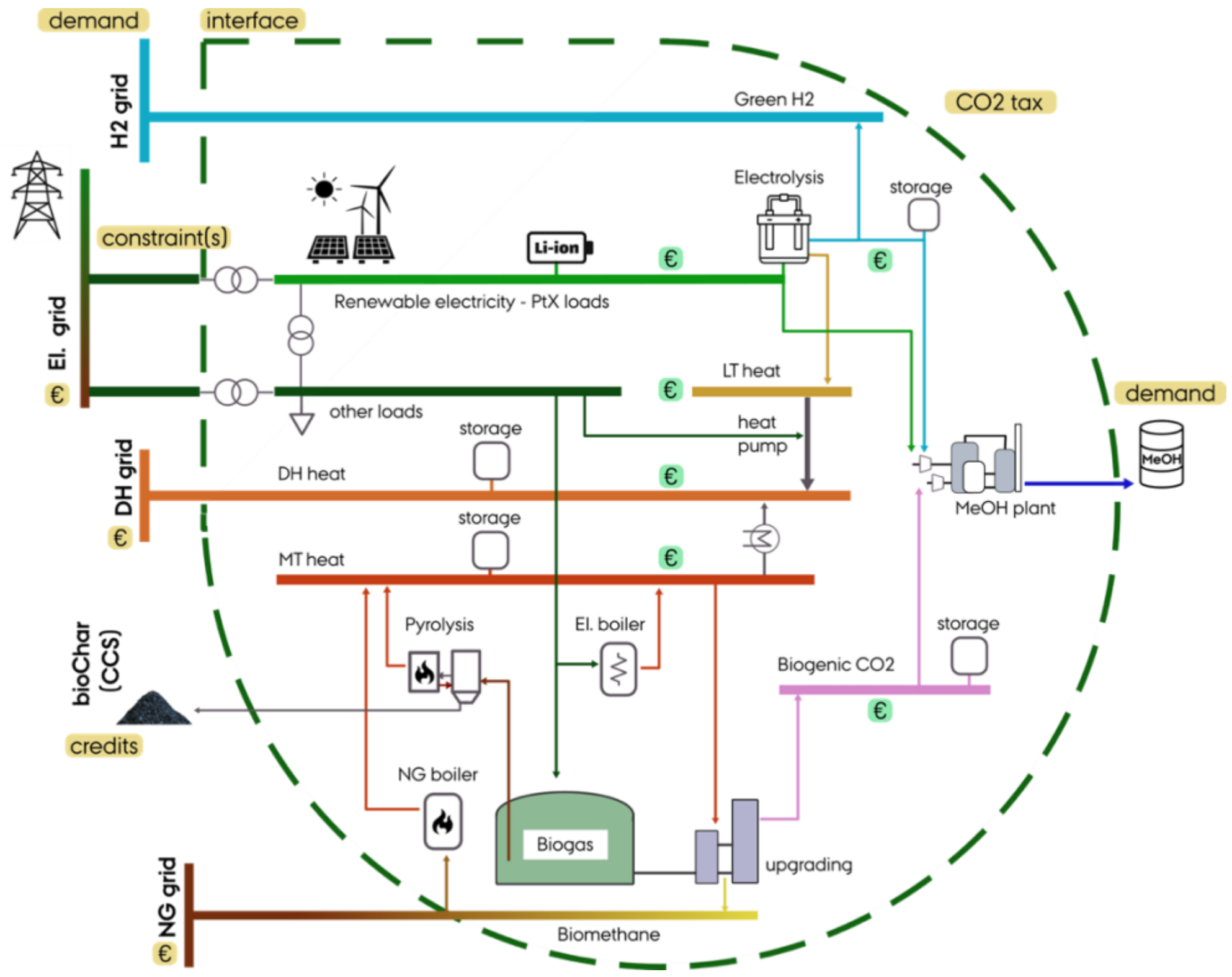

**Figure 1**: Overview of the PtX hub, with indication of the optimization boundaries and an example of the technology options (existing and potential).

Many other projects are arising in the EU (Table.2) and the capacity of individual plants is expected to rise form 5-10 kt/y to 50-25 kt/y within the next decade [37]. Large-scale plants will be based on CO2 captured form incineration of municipal solid waste or biomass fired plants. Recent advancements in electric steam methane reforming [11] enabled coupling of SMR with renewable electricity and reduced the volumetric size of the reactors making biogas a suitable feedstock for convectional methanol production. Nevertheless, this process is not considered in this study due to the high forecasted demand for biomethane in the upcoming years.

## 2.3 Legal framework

The EU's framework is grounded in several key regulations and directives, notably the Renewable Energy Directive II [3] (RED), which sets the targets for the adoption of renewable energy across sectors. For hydrogen and RFNBOs, the focus is on ensuring that these fuels are produced sustainably and contribute to the overall reduction of greenhouse gas emissions. The framework incorporates three fundamental principles to ensure the integrity and sustainability of hydrogen and RFNBO production: additionality, temporal correlation, and geographic correlation. Additionality requires that the renewable electricity used to produce hydrogen and RFNBOs must come from new renewable energy projects that are developed specifically for this purpose. The production of hydrogen and RFNBOs should not divert existing renewable energy from other essential uses, thereby driving the creation of additional renewable energy capacity and contributing to the overall increase in renewable energy supply. Temporal correlation mandates that the renewable electricity used must be consumed simultaneously or within a specific time frame relative to its generation. This principle guarantees that the hydrogen and RFNBO production is directly linked to the availability of renewable energy, enhancing promoting grid stability. However until 1st January 2030 temporal matching will occur monthly rather than an hourly one. Geographic correlation stipulates that the renewable energy used for hydrogen and RFNBO production must be generated within the same geographic area where it is consumed, supporting regional energy security. The Renewable Energy Directive mandates that synthetic fuels, classified as renewable liquid and gaseous fuels of non-biological origin (RFNBOs), must exhibit a substantial 70% reduction in greenhouse gas (GHG) emissions throughout their entire life cycle. Consequently, e-fuels distributed and endorsed in Europe within the existing regulatory framework do not guarantee a complete 100% emissions reduction, requisite for attaining $CO_2$-neutrality. This is due to the potential utilization of non-sustainable $CO_2$ sources and grid electricity in e-fuel production.
Furthermore, the RED permits the utilization of $CO_2$ obtained from industrial sources falling under the Emissions Trading System (ETS) and presently excludes carbon captured from the atmosphere via Direct Air Capture (DAC). The utilization of industrial emissions is permissible until 2035 for emissions stemming from power generation and until 2040 for other ETS-covered sources.

While these timelines are established within the context of an economy striving for carbon neutrality by 2050, they remain subject to periodic reassessment, contingent on the progress and compliance within the sectors outlined in the Directive (2003/87/EC) with respect to the 2040 targets. The discussion regarding the legal definition of renewable e-fuels lies at the crossroads of several interests and is likely to remain a debated topic in the years to come. This debate is heavily influenced by the gap between the substantial demand for fossil fuels (including the transport sector) and the challenges in upgrading energy systems to supply their renewable counterparts at a competitive cost. Another pertinent factor is the parallel discussion on EU carbon tax, which influences markets for fossil fuels and could offset the cost penalty associated with transitioning from fossil fuels to e-fuels and biofuels.

After 2030, the EU aims to strengthen its $CO_2$ tax framework, with higher tax rates and more rigorous enforcement which will affect all sectors. Future prices of EU ETS credits remain uncertain, the overall trend points towards significant increases due to tightening emission caps, heightened climate policies, and ongoing economic and technological developments.
Current $CO_2$ prices have been relatively stable within the € 90-€ 100 range on the ETS market. However, given the EU’s commitment to net-zero emissions by 2050, ETS credit prices are expected to continue rising post-2030, potentially reaching €150-€ 250 per ton or higher as we approach 2050. For the scope of this investigation, we set an overall $CO_2$ tax for fossil emission and carbon capture credits at 150 (€/t$CO_2$) in the base case and 250 (€/t$CO_2$) for sensitivity analysis (table 1), in addition to a case without $CO_2$ tax to investigate the trend.

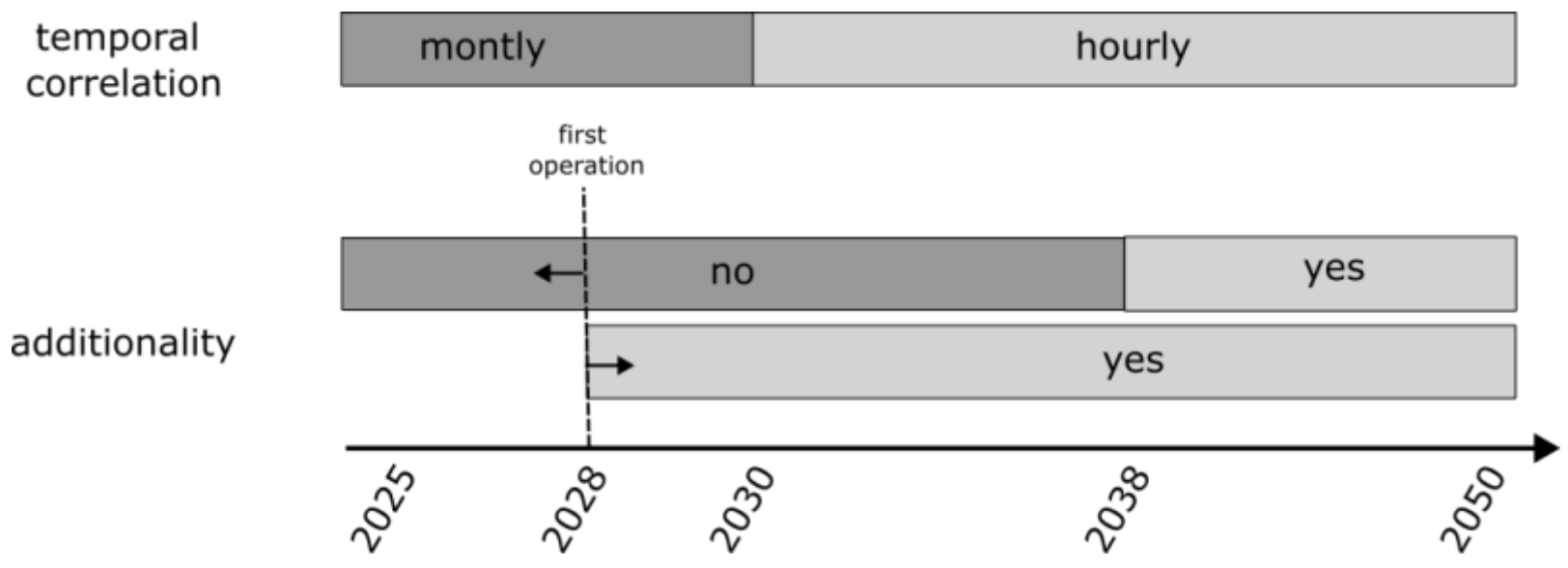

**Figure 2** – roll-out of RFNBOs directive

## 2.4 Overview of GreenLab Skive PtX hub concept

The GreenLab Skive PtX hub concept is based on the integration of several separated stakeholders within in industrial symbiosis. The hub is grounded around a modern large-scale biogas plant and represent an expansion model replicable in multiple location within the EU. The Biogas plant collects manure, waste and energy corps for the local surrounding and produces biomethane and releases biogenic $CO_2$ to the atmosphere. The PtX hub hosts several industries with a variety of application related to energy production and non as: on-site wind and solar farm, hydrogen and e-fuels, plastic recycling, green protein production, biomass pyrolysis, among the others. These separate stakeholders trade energy and material flows potentially benefitting from lower prices in the behind-the-meter market created at GLS. Only the stakeholders with business related to the PtX operation were considered here (Figure 1). Although, the actual GLS site deviates from this model in terms variety of internally traded flows, we consider this representation to be more general and valuable for the purposes of the investigation. The stakeholders included in the optimization can be briefly categorized in six groups, which are further used in the model. The *Skive Biogas* plant specializes in biomethane production for the grid. It boasts a production around of 200 GWh of biomethane per year and employs amine scrubbing for the $CO_2$ removal process. Additionally, a natural gas boiler is utilized to supply the necessary heat for amine regeneration. The *$H_2$ production* refers to the alkaline electrolysis plant that delivers hydrogen at 35 bar, for use at the PtX hub or injection in the $H_2$ grid. *MeOH production* pertains to the methanol synthesis plant, which operates through the hydrogenation of $CO_2$ obtained from biomethane upgrading. This plant incorporates processes for $CO_2$ and $H_2$ compression and short-term storage of both gases.

The *Central Heating* includes those technologies that supply heat to the network in the PtX hub. They included additional electric or natural gas boilers and pyrolysis of biogas digestate residues (utilizing SkyClean technology) with the added benefit of negative $CO_2$ emissions from biochar storage. *Renewables* denotes the onshore wind and solar photovoltaic power generation that can be installed on-site or purchased through Power Purchase Agreements (PPAs) within the same vicinity. The final stakeholder, referred to as *Symbiosis Net,* acts as the local site operator, investing in and managing the infrastructure necessary for internal trading of renewable electricity, hydrogen, carbon dioxide, and heat at different temperatures. The *Symbiosis Net* provides the essential infrastructure for transmission (such as cables, transformers, or piping), intermediate storage (utilizing lithium-ion batteries and thermal energy storage), and interfaces with external grids for electricity (through transformers), natural gas and district heating (using heat pumps for waste heat upgrading).

## 2.5 Technology inputs and costa data

The linear optimization model requires several inputs data in particular cost estimates and performance of all plants and key technologies present in the PtX hub.
Whether, possible the technology data catalogue from the Danish Energy Agency (DEA) [38] was used as reference for both technology and cost inputs to the model, other sources where introduced if necessary. Here, follows a description of the technology available for each stakeholder and the input and cost data used in the study. A deeper review of technologies peculiar to GLS and present in the model is available in Appendix B. Other common energy technologies as wind turbine, solar PV, gas fired and electrical boilers are not further described in this study, but refer to the technology data catalogue [38]. All the cost inputs to the optimization model are summarized in table 4 and technical performance parameters in tables 5 and 6.

**Skive Biogas**

The biogas plant is designed to process up to 500 000 tons per year of animal manure and residual products from agriculture and food industry. The plant is usually operated to process 300 000 tons per year sources in the neighbor regions corresponding to a production of about 200 GWh of biomethane. Biogas is upgraded to biomethane trough an amine scrubber which required 0.085 MW of heat for each MW of biomethane currently obtained through combustion of natural gas. The product biomethane is compressed to the grid pressure of 40 bar for injection. As aforementioned, no capital costs are associated to the standard operation of the Skive Biogas plant and all the investments, including the NG boiler are considered sunk costs. The energy demands to operate the complete biomethane plant are reported in table 5, while a more detailed description of the process and the mass and energy balance is available in appendix B.

**Electrolysis and hydrogen grid connection**

In this investigation only alkaline electrolysis cells are considered in this study due to their cost advantage and increasingly high availability on the market. Other technologies as polymer electrolyte membrane and solid oxides cells are not included. The process yields hydrogen, oxygen, and surplus heat as output. While oxygen is presently not utilized in GLS, there exists potential for its application in various processes. The hydrogen-to-electricity efficiency of AEC systems is projected to be 62.2% by 2030, based on the lower heating value (LHV). The excess heat, emanating from the process, is dissipated at around 50 °C. According to the DEA, approximately 22.3% of the electricity generated is converted into heat suitable for recovery to district heating. However, due to GLS's considerable distance from urban settlements, the district heating temperature in this study was set at 90 °C. Consequently, none of the excess heat from electrolysis is available for district heating without upgrading via a heat pump. The hydrogen output pressure is assumed to be 35 bars equal to the pressure for injection in the distribution grid and for local up-takers. The cost of hydrogen-gird infrastructure and compression from DSO to TSO level are not included. The electrolysis plant has ramp-up time below 1h from zero to maximum capacity in cold conditions and few minutes in normal operation. The forecasted investment cost for 2030 is set to 575 (€/$kW_{el}$) for a large plant (e.g.100 MW) which includes all components required for converting 400VAC electricity and purified water into $H_2$ gas at 35 bar and a waste heat stream at 50 °C. For smaller plants (e.g. 10 MW) the investment cost increases to 900 (€/$kW_{el}$) and is used for stand-alone MeOH production. All the techno-economic inputs for alkaline electrolysis refer to the latest version of the DEA catalogue (2024) [19]. Other technical data are available in table B. In the study cases where hydrogen is injected to the grid it is assumed that the pressure at the distribution grid is lower than 35 bar and there is not a final compression stage at the PtX site.

**E-methanol synthesis**

Methanol synthesis at GLS is based on $CO_2$ hydrogenation process. Several leading companies are licensing direct $CO_2$ hydrogenation processes using commercial selective catalysts are based on $Cu/ZnO/Al_2O_3$. Research and development is underway to increase methanol selectivity (hindered by significant reverse water−gas shift reaction) and improve stability undermined by water-induced sintering of the active phase [39] [40]. The technology data catalogue from the Danish Energy agency for $CO_2$ hydrogenation is based on commercial processes, in particular the George Olah Renewable Methanol Plant [26] located in Iceland and operated since 2011. The process schematics for e-methanol production addressed in this study are reported in appendix B, the process is adapted from the DEA technology data catalogue [41]. The process is broadly divided in a MeOH synthesis section and a distillation section. In the synthesis section hydrogen and carbon dioxide are fed at 65 bars with a molar ratio of 3. The two streams are mixed, compressed to 80 bars and pre-heated before entering the synthesis reactor. The whole process is integrated to recover the reaction heat from the exothermic MeOH synthesis to pre-heat the process streams and distillation.
Cost inputs and mass and energy balances from the technology data catalogue have been adapted with data from other sources to account for compressors and storages [19] [20] [42] [14]. The main inputs for the process are reported in Tables 4 and 5, for a more detailed description see table B.2 in appendix. To represent this a ramp-up and -down constraints are set in the model equal to 48h for increasing production from zero to full capacity, or vice versa. Methanol synthesis is not a process which offers flexibility towards intermittent operation as temperature and pressure conditions must be maintained throughout the whole process for efficient and safe operation. Hence, storage of $CO_2$ and/or $H_2$ is required for coupling with intermittent energy sources. Capacity and operation of hydrogen and $CO_2$ compressors and storages are optimized separately from the MeOH synthesis process, although they are conceptually part of the methanol plant.

**$H_2$ and $CO_2$ compression and storage**
The hydrogen used for MeOH synthesis is compressed from the electrolyzer pressure of 35 bar to 60 bar before mixing with carbon dioxide and further compression at the MeOH plant. Hydrogen storage is possible with a storage tank before the mixing with carbon dioxide. Compression of carbon dioxide occurs from atmospheric pressure to 60 bars for mixing with hydrogen. Here we assume that maximum temperature of the gases out of each compression stage is 130 °C and the heat in the intercooling stages can be recovered in the heat network connected to the district heating (130-80 °C). Two options for storage of $CO_2$ are available in the model (Fig 5.3) with different characteristics for capacity, energy losses and duration: 1) high pressure storage at 60 bars and ambient temperature through a system of interconnected cylinders for a capacity up to 1 ton (about 40 cylinders), 2) bulk storage of liquefied $CO_2$ in insulated tank at 16 bars. Technical characteristics and costs for the different storage options are reported in Tabs 4 and 5, the electrical consumption associated with the use of each storage option was reported as difference compared to the steady state operation of the MeOH process. More information is available in appendix B. Here we assume that the storage capacity for hydrogen is not limited, although legal permits assigned to the energy site may limit the maximum amount of hydrogen stored on the site.

**Centralized Heating Technologies**
These are conventional heat generation technologies able to provide heat to the MT heating network in addition to the pyrolysis plant the existing NG boiler and the excess heat from the industrial processes. The technologies included here are: additional NG boiler, electrical boiler, biomass boiler using pellets from digestate fibers. Connection of each plant to the heat network has an added cost for the heat exchanger. All the techno-economic assumptions refer to the DEA catalogue.

**SkyClean pyrolysis plant**
The biogas plant is coupled to the Stiesdal SkyClean pyrolysis plant. The main feedstock to the pyrolysis plant are the fibers obtained from the solid fraction of the biogas digestate which are dried, pelletized and converted in an updraft pyrolysis reactor optimized to produce biochar with high stability for carbon storage [43]. The other pyrolysis products are combusted in a gas boiler producing heat for the biomethane upgrading (figure B.2 and B.3). The biochar produced at SkyClean is undergoing the process for approval as carbon storage technology, in this work we assumed that it can be considered a negative emission technology (see appendix B.3 for more details).
**Table 4**: main cost assumptions for year 2030.

| Technology | Reference flow | Investment cost | Fixed O&M | Variable O&M | Lifetime | Source |
|---|---|---|---|---|---|---|
| | | (k€/ ref.) | (%/y) | (€/$MWh_{ref}$) | (y) | |
| | | | **Renewables** | | | |
| On-shore wind | El. (MW) | 1040 | 1.22 | 1.35 | 30 | [41] |
| Solar PV | El. (MW) | 380 | 1.95 | 0 | 40 | [41] |
| Grid connection | El. (MW) | 140 | 2 | - | 40 | [41] |
| | | | **Hydrogen production** | | | |
| Alkaline electrolysis (100 MW) | El. (MW) | 575 | 4 | - | 25 | [19] |
| Alkaline electrolysis (10 MW) | El. (MW) | 900 | 4 | - | 25 | [19] |
| Water purification | Water (t/h) | 135 | 2 | - | 25 | internal |
| | | | **Centralized heat production** | | | |
| SkyClean (inc. drying) | Dry pellets ($MW_{dry}$) | 2011 | 3.64 | 11.6 | 25 | [44] |
| NG boiler | Heat (MW) | 50 | 1.04 | 1 | 20 | [44] |
| Electric Boiler | Heat (MW) | 70 | 1.0 | 1.0 | | [44] |
| Biomass boiler | Heat (MW) | 590 | 7.5 | - | 20 | [41] |
| | | | **E-methanol** | | | |
| Synthesis process | MeOH (MW) | 651 | 3.0 | - | 30 | [41] |
| $CO_2$ compressor | $CO_2$ (t/h) | 1516 | 4.0 | - | 15 | [19] [41] |
| $H_2$ compressor | $H_2$ (MW) | 79.4 | 4.0 | - | 15 | [41] |
| $CO_2$ liquefaction & evaporator | $CO_2$ (t/h) | 19.76 | 5.0 | - | 25 | [41] & internal |
| $CO_2$ storage (liquid tank) | $CO_2$ (t) | 2.53 | 1.0 | - | 25 | [45, 46] |
| $CO_2$ storage (cylinders) | $CO_2$ (t) | 77.2 | 1.0 | - | 25 | internal |
| $H_2$ storage | $H_2$ (MWh) | 12.3 | 2.0 | - | 20 | [41] |
| | | | **Symbiosis Net** | | | |
| Transformers & grid connection | El. (MW) | 0.140 | 2.0 | - | 40 | [41] |
| Li-ion battery | El. (MWh) | 0.142 | - | - | 25 | [47] |
| Battery inverter | El. (MW) | 0.160 | 0.34 | - | 10 | [47] |
| Hot water tank | Heat (MWh) | 0.540 | 0.55 | - | 25 | [47] |
| Thermal battery | Heat (MWh) | 25 | - | - | - | |
| Local Heat network | Heat·km (MW·km) | 25 | - | - | - | [48] |
| Heat exchangers | Heat | 100 | - | - | - | [48] |
| Local $H_2$ pipe | $H_2$·km (MW·km) | 3.8 | 3.17 | - | 50 | [48] |
| Local $CO_2$ pipe | $CO_2$·km (t/h·km) | 130 | 0.1 | - | 50 | [38] |
| Heat pumps | Heat (MW) | 780 | 0.11 | 3.2 | 20 | [44] |

The actual Skyclean plant at GLS is pilot of 2MW dry pellets input, however in this study the maximum capacity of the plat is set to 40 MW as planned for the largest commercial plant of this type [41]. Beside we assumed that pellets can be purchased on the market at the price of 380 €/t to increase the capacity. The main product of the SkyClean technology is medium temperature heat supplied to the amine scrubber and other customers in the site, together with the stable biochar for carbon sequestration which is produced with a yield of equivalent to 0.173 ($tCO_2/MWh_{pell}$) The energy and mass balance of the pyrolysis system are reported in table 5, the energy demand includes dewatering of digestate fibers, drying and pelletization. The SkyClean system is considered internally heat integrated covering the demand for drying. It is worth nothing that in this study the opportunity cost for alternative use of the fibers is set to zero, although an option for use in a biomass boiler is considered among the heating technologies.

**Symbiosis net**

The symbiosis net at GLS includes a set of technologies which enable the internal trading of energy and material flows between the different plants, e.g. electricity, heat at different temperatures, carbon dioxide, hydrogen, etc… The costs and technical assumptions related to this infrastructure refer as much as possible to the DEA database. The cost for the electrical infrastructure is roughly estimated by the cost of transformers and point of connection to the external distribution grid (60kV). The three heat network possible in the GLS model are interconnected allowing for heat integration (Figure B.7). Storages of heat is possible with two technologies: a thermal battery operating at temperatures as high as 350 °C using electricity and a hot water tank for storage at about 95 °C. Input data for the thermal battery refer to commercial concrete-based technology [49]. The thermal battery system can be designed with an heat exchanger for waste heat recovery and/or with electric resistance for electrical heating. In the optimization, the actual coupling between high temperature heat source and the thermal battery is not analyzed and the thermal battery is just placed on the MT heat network. With this simplification any high temperature heat source in the model (NG boiler, electric boiler, pyrolysis plant op biomass plant) can charge the thermal battery, although the design of the system will be different depending on the heating technology. The cost for the heat network infrastructure was estimated using an average piping length of km and a cost for all the heat exchangers was estimated based on the DEA database [48]. An optional heat pump system can be installed to upgrade LT heat, mostly from the electrolysis plant up to 90 °C. The sale of district heating is accounted with a state-controlled price of 54 €/$MW_{th}$ [50]. Further details about cost and performance are reported in appendix B.

**Table 5** – Main technology data assumptions for 2030 – as model inputs.

| | **Biomethane plant** | **SkyClean** | **Electrolysis** | **Methanol Synthesis** | **$CO_2$ compression** | **$H_2$ compression** | **Heat Pump** |
|---|---|---|---|---|---|---|---|
| Reference flow | bioCH4 | Dry pellets | Electricity | Methanol | $CO_2$ (1-80 bar) | $H_2$ (35-80 bar) | Electricity |
| | | | **Inputs** | | | | |
| Heat input 180°C> T >150°C | 0.103 ($MW_{th}/MW_{ref}$) | | | 0.105 ($MW_{th}/MW_{meoh}$) | | | |
| Heat input 80°C> T >50°C | | | | | | | 1.7 ($MW_{th}/MW_{el}$) |
| Electricity | 0.04 ($MW_{el}/MW_{CH4}$) | 0.067 ($MW_{el}/MW_{pell}$) | | 0.018 *** ($MW_{el}/MW_{meoh}$) | 0.096 ($MWh/t_{CO2}$) | 0.010 ($MW/MW_{H2}$) | |
| $H_2$ | | | | 1.155 ($MW_{H2}/MW_{meoh}$) | | | |
| $CO_2$ | | | | 0.253 ($t/h_{CO2}/MW_{meoh}$) | | | |
| | | | **Outputs** | | | | |
| $H_2$ | | | 0.622 ($MW_{H2}/MW_{el}$) | | | | |
| $CO_2$ | 0.0982* ($t/h_{CO2}/MW_{CH4}$) | | | | | | |
| $CO_2$ emissions | | -0.164** ($tCO_2/MWh_{pell}$) | | | | | |
| Digestate Pellets* | 0.09 ($MW_{pell}/MW_{CH4}$) | | | | | | |
| Heat output T>150°C | | 0.36 ($MW_{th}/MW_{pell}$) | | | | | |
| District Heating output 150°C>T>90°C | | | | 0.256 ($MW_{th}/MW_{meoh}$) | | | 2.7 ($MW_{th}/MW_{el}$) |
| Heat output 90°C>T>50°C | | | 0.223 ($MW_{th}/MW_{el}$) | | | | |
| | | | **Constraints** | | | | |
| Minimum load | | | 15% | 20 % | | | |
| Ramp up/down time (0%-100% load) | - | 12 h | < 30 min | 48 h | | | $\leq$ 1h |
| Source | [41] | [44] | [19] | [19, 41] | [19, 21, 51] [52] | [52] [21] [45] | |
| [*]biogenic, [**]$CO_2$ equivalent of sequestered biochar, [***] without compression of $H_2$ and $CO_2$ | | | | | | | |

**Table 6:** Technology data assumptions for 2030 – as model inputs.

| | **$CO_2$ cylinders storage** | **$CO_2$ liquefaction and tank storage** | **$H_2$ tank Storage** | **Thermal Energy Storage** | **Thermal battery (concrete)** | **Lithium-ion battery** |
|---|---|---|---|---|---|---|
| Reference flow | $CO_2$ (t) | $CO_2$ (t) | $H_2$ (t) | (Water 95 ℃) | Heat (MWh) | Electricity |
| **Inputs** | | | | | | |
| Pressure | 60 bara | 16 bara | 60 bara | | | |
| Temperature | <50℃ | - 27 ℃ | < 120℃ | < 100℃ | 350 – 150 ℃ | |
| Electricity* (inc. extra compression) | | 0.077 (MWh/$t_{CO2}$) | | | | |
| **Constraints** | | | | | | |
| Capacity (min-max) | | | | | | |
| Charging rate/capacity | | $\geq$ 1 & $\leq$ 15 ($t_{CO2}$/h) | | $\leq$ 1/6 ($h^{-1}$) | $\leq$ 1/6 ($h^{-1}$) | $\leq$ 1 ($h^{-1}$) |
| Discharge rate/capacity (min-max) | | $\geq$ 1 & $\leq$ 15 ($t_{CO2}$/h) | | $\leq$ 1/6 ($h^{-1}$) | $\leq$ 1/6 ($h^{-1}$) | $\leq$ 1 ($h^{-1}$) |
| Source | | [45] | | [41] | [49] [53] [54] | [47] |
| * Increase from steady-state MeOH process | | | | | | |

# 3 Methodology

The analysis of cost for hydrogen and e-methanol production within the integrated PtX hub utilized the PyPSA (*Python for Power Systems Analysis*) linear optimization model, as referenced in [55]. Appendix A provides a comprehensive outline of the mathematical framework underlying the PyPSA model of GLS. In this investigation, PyPSA was deployed to identify the combinaton of generation, conversion, and storage technologies that minimizes the total system cost while ensuring a consistent supply of energy and materials flows to meet demands. The optimization variables encompass the capacities and power flows of each technology, computed with a 1-hour time resolution. During optimization, we specified the annual demands for hydrogen and methanol, alongside the fixed demand and capacity for the biomethane plant.

The model operates on the premise of an efficient market characterized by perfect competition among all featured technologies and the maintenance of enduring market equilibrium, ensuring that all energy technologies fully recover their costs. The capacity of all plants and components is left unconstrained in the formulation of optimization problem, except for the original biomethane plant, which serves as a benchmark. No upfront investment cost is attributed to the biomethane plant, and only differeces in operational cost considered. At each time step a shadow price, also referred to as the Karush-Kuhn-Tucker (KKT) multiplier, associate with the demand constraint, indicates the marginal price of the energy carriers within the system. In this context, the KKT multipliers for the demand constraints of hydrogen and methanol were interpreted as levelized cost for production of these carriers. Meanwhile, shadow prices for internal buses indicate the equilibrium prices for behind-the-meter trading at the PtX hub, such as the local marginal price for internal electricity or $CO_2$ buses. Furthermore, the model assumes a level of foresight capable of accurately anticipating energy supply and demand throughout the year, incorporating year-ahead weather forecasts, energy prices, and emission intensities of external energy systems. Consequently, any form of storage intended as a safeguard against unforeseen energy shortages is not considered into this model.

The interactionbetween the PtX hub and the external energy systems is modelled though and interface, facilitating the trade of energy commodities at exogenously determined prices. Historical prices from 2019 and 2022 were utilized for electricity and natural gas, representing scenarios with low and high energy prices. The interface enables the import or export of electricity, natural gas, district heating, and biochar storage, with prices determined by historical spot prices, tariffs, and $CO_2$ taxes. The spot prices at the interface are assumed to be inelastic, disregarding supply or demand responses related to the PtX hub's operation. Two constraints are imposed at the interface: one restricts the use of grid electricity for hydrogen and e-methanol production in accordance with RFNBO legislation for investments beginning in 2030, while the other limits the maximum sales of renewable energy to the external grid to prevent the optimal solution from excessively favoring electricity trading, thus altering the core business of the PtX hub.

All costs and estimated lifespans of technologies utilized in the model are documented in Table 4, referring to the year 2030 to provide reliable cost estimates for the forthcoming development of the hydrogen and e-fuel economy. Investment expenses are amortized over the reported lifespan of each asset, considering a discount rate of 7%. The optimization model was applied to two different scenarios (Table 1), and the results were derived from parametric analysis across a set of key parameters (Table 8). Instead of modifying cost assumptions, the focus was placed on design parameters, the impact of external market conditions (such as energy prices, $CO_2$ taxes, and the amount of electricity that can be sold to the external grid), and additional technologies like district heating and biochar storage with carbon credits.

## 3.1 PyPSA model

Figure 3 provides a simplified representation of the model, depicting the main buses, generators, links, and storage units representing the available technologies. To maintain simplicity in the figure, some intermediate buses and other components have been omitted. A schematic with a higher level of detail is available in the appendix E, for the scenario *$H_2$ to Grid* before optimization of the network and more schematics can be obtained in the Github repository. The demands for hydrogen and methanol, which drive the optimization, were set annually affording full flexibility to optimize operation for the lowest cost. All the energy conversion processes were modeled using the PyPSA multilink component, based on assumptions detailed in appendix B. Renewable energy generation relies on generators with time-dependent availabilities ($\underline{g}_{n,s,t}$ and $\overline{g}_{n,s,t}$), matching the capacity factors for onshore wind and solar in Skive (Appendix D). On the left-hand side of Figure 3, the buses forming part of the external grids (DK1, NG, DH and bioChar) are displayed. Within the PtX Hub, two electricity buses have been identified. The first (El3) connects RE generation and hydrogen production, while the second (El2) links all other loads, which do not necessarily require RE sources. These two electricity buses are connected by a link (RE to hub) and are independently linked to the external grid. The model of the heat network in GLS can operate on three temperature levels (Figure B.7): 1) pressurized hot water with temperature in the range 180-150 °C (MT heat), 2) pressurized hot water for temperature 150 - 90 °C (DH heat), and 3) hot water in between 90 and 50 °C (LT heat). The three networks satisfy specific demands the medium temperature network supplies heat to the amine scrubber and the methanol distillation, the district heating network collects waste heat above DH temperature (90 °C) for sale to DH and the low temperature network collects waste heat below DH temperature for upgrading via heat pumps.

The interface between the internal GL market ant the external grids are represented by the following links: *DK1 to El3* for purchasing of grid electricity, *EL3 to DK1* for sales of RE, *GLS to DH grid* for sales of DH and *biochar credits* for sales of biochar credits. Exogenous prices are applied to these links. The constraint the purchasing of grid electricity for RFNBOs was set on the link *DK1 to El3* limiting purchasing only to those hours with a spot price below 20 €/MWh (Eq. 6). The constrain limiting the sales of RE to the grid was not applied to the capacity of the *EL3 to DK1* link, but modelling the external electricity demand such that it follows the profile of the total demand on the DK1 grid and it was scaled accordingly (eq. 7). This demand could be supplied either via RE at the PtX hub (with revenues from sales) or via a zero-cost generator. This approach guaranteed that the objective function included only the cost and revenues of the PtX hub, and sales of RE could be optimized independently form the external grid demand. The scaling of the external electricity demand if controlled by the parameter *maxRE* which is the ratio of the maximum annual RE sellable to the grid and the energy used for PtX processes (both $H_2$ and MeOH depending on the scenario). Hence, it sets a balance between the core business of the park as PtX hub and side business as power producer. The parameter *maxRE* is part of the sensitivity analysis, ranging from a minimum of 0.1 where the grid is used only for balancing the RFNBOs production to a maximum of 1 where both businesses are equally relevant. The external district heating grid is modeled similarly to the electricity using a zero-cost generator to satisfy the exogenously set demand, which was obtained for the Skive municipality (appendix C) and fixed average price corresponding to 400

DKK/MWh. Biochar credits were rewarded at the same value of the $CO_2$ tax. The prices for purchasing and sales of electricity and for NG were calculated from the historical spot prices [56] and updated $CO_2$ taxes and tariffs for TSO [57] and DSO [58]. The calculations are shown in tab 7. The total cost for electricity purchased from the grid ($p_{el,t}^{in}$) is calculated as the spot price for the reference year ($p_{el,t}^{ref}$) adjusted for the difference in $CO_2$ tax between the reference year ($p_{CO2}^{ref}$) and the scenario considered ($p_{CO2}^{SC}$), then summed with the total electricity purchase tariffs for 2025 ($TF_p$), showed in Tab 7. For sales of renewable electricity to the grid, the price is set equal to the electricity spot price minus the tariffs for electricity sale [56]. The prices for electricity within PtX hub, represented by the shadow prices at the El2 and El3 buses, is a combination of the electricity price from the grid and the levelized cost of energy (LCoE) for renewables on site. Hence, if the prices are such that the LCoE for RE is lower than the purchase price of grid electricity, then the shadow prices will fall within in the window defined by the prices for purchase and sale of grid electricity, indicating that behind-the-meter trading is profitable for companies within the park.

As methodological assumption the sales of biomethane were not included in the objective function, the capacity and operation of the biomethane plant were fixed and the cost for standard operation of the reference biomethane process was set equal to zero. The standard operation includes the original consumption for NG and electricity compunction from the grid with adjusted priced, hence was independent from the scenario. As results of this assumption the shadow price for the biomethane bus represents a difference in production cost of biomethane compared to the reference operation (can be negative) due to variation the cost for electricity and heat in the integrated hub.

**Table 7** - Interface between external grids and behind-the-meter market

| | | |
|---|---|---|
| Price for purchase of electricity from the grid | $p_{el,t}^{in} = p_{el,t}^{ref} + TF_p + em_t \cdot p_{CO2}^{SC}$ | (1) |
| Price for sales of electricity to the grid | $p_{el,t}^{out} = p_{el,t}^{ref} - TF_{sl}$ | (2) |
| Price for purchase of NG from the grid | $p_{NG,t}^{in} = p_{NG,t}^{ref} + em_{NG} \cdot p_{CO2}^{SC}$ | (3) |
| Price for sales of DH to the grid | $p_{DH} = 54\ (\frac{€}{MWh_{th}})$ | (4) |
| Price for sales of biochar credits | $p_{bioChar} = p_{CO2}^{SC}(\frac{€}{t_{CO2}})$ | (5) |
| Constraint for use of grid electricity for RFNBOs | $\bar{f}_{gridtoH2,t} = 1 \quad \forall\ t,\ \ el_{price,t} \leq 20\ \left(\frac{€}{MWh}\right)$ | (6) |
| Constraint on maximum sales of electricity | $\sum_t d_{elDK1,t} = \left(\frac{\sum_t d_{H2,t}}{\alpha_{H2/el}} + \frac{\sum_t d_{MeOH,t}}{\alpha_{MeOH/el}}\right) \cdot maxRE$ | (7) |
| * $\alpha_{MeOH/el}$ includes the average consumption for compressors | | |

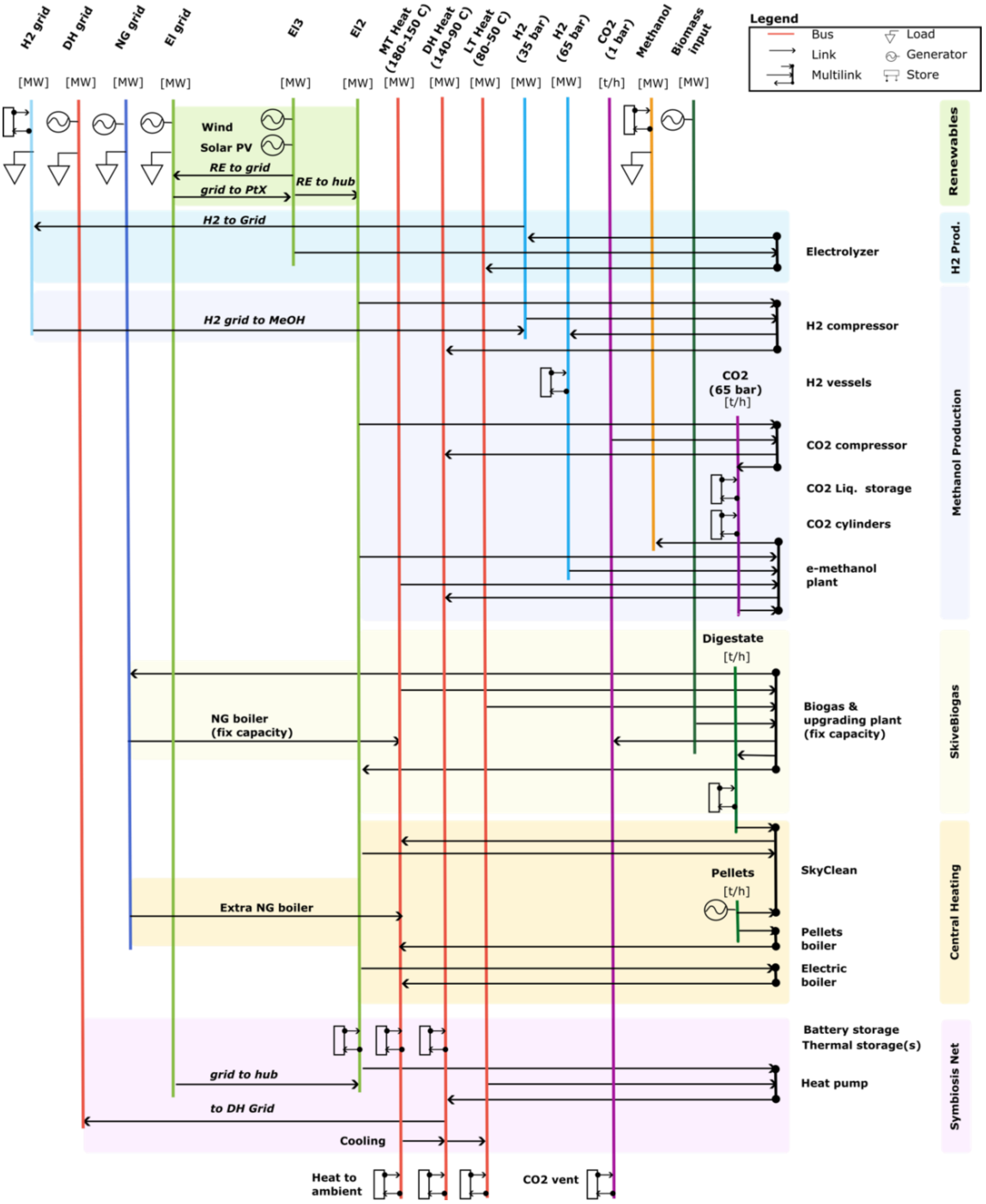

**Figure 3**: Generic superstructure of the PyPSA model (*$H_2$ to Grid* scenario), showings the technologies portfolio and potential energy and material integration.

## 3.2 Sensitivity analysis

The sensitivity analysis covered several aspects of PtX hub, which are summarized in Table 8. It was carried by optimizing the PyPSA network for all the combinations of the six sensitivity parameters and for each of the two scenarios *H2 to Grid* and *MeOH standalone*. The sensitivity selected are: 1) Energy year price (2019 low prices and 2022 high prices), 2) ratio between maximum RE sales and electricity for RFNBOs, 3) ratio of $CO_2$ from biomethane recovered to MeOH (which sets the MeOH demand), 4) $CO_2$ tax for fossil emission or biochar credits, 5) DH enabled in the model, 6) biochar credits enabled in the model.

**Table 8 :** sensitivity parameters

| Parameter name | Type of parameter | Range | Reference value |
|---|---|---|---|
| Hydrogen demand to the grid | Binary (compares standalone MeOH and $H_2$ for Grid connection ) | True, False | - |
| $CO_2$ tax | Integer (affects marginal costs) | 0, 150, 250 (€/t) | 150 |
| Ratio of $CO_2$ converted to MeOH | Integer (sets a demand ) | 0.8, 0.85, 0.90, 0.95, 0.99 | 0.9 |
| Ratio between max electricity sold to the grid and electricity for PtX | Integer (sets a constraint) | 0.1, 0.5, 1 | 0.5 |
| Energy price year | Lumped inputs (El. prices, NG price, El. emissions intensity) | 2019, 2022 | - |
| District heating | Binary (allows for DH conenction) | True, False | False |
| Biochar credits | Binary (allows for revenues from biochar) | True, False | False |

# 4 Results and discussion

The results are presented using as reference an arbitrary set of sensitivity parameters and variation from it are discussed along with the results. The reference set of parameters is indicated in table 8 and does not include the $H_2$ demand to grid and energy year price which are always explicitly indicated in the figures.

## 4.1 Production cost of $H_2$ and MeOH

The estimated production cost of $H_2$ and MeOH is shown in figure 4 as function of the two parameters that exhibit the highest sensitivity: the ratio of $CO_2$ recovered to MeOH, and the ratio of electricity sold to the grid. The uncertainty bands in each figure represent the variability to the other parameters. The production cost for $H_2$ is estimated in 87-92 €/MWh for a grid connected plant and 107-111 €/MWh for supply to standalone MeOH production. The H2 production cost for the *$H_2$ to grid* cases is in line with a forecasted price of 3 €/$kg_{H2}$, but they would not be competitive in a 2 €/$kg_{H2}$ scenario.
The MeOH production cost is considerably affected by the rate of $CO_2$ recovery to MeOH, especially for *MeOH standalone* cases. Achieving a $CO_2$ recovery ratio beyond 90% required higher complexity in the plant with large storage capacity for $CO_2$ and $H_2$, increasing the investment cost. For a 90% recovery rate the MeOH production cost is estimate in 111-114 €/MWh for the *$H_2$ to grid* cases and 135-142 €/MWh for the MeOH standalone cases. For comparison fossil methanol (priced at 360 €/t) would require a $CO_2$ tax of 180 (€/$t_{CO2}$) to reach a similar market price. Therefore, the coupling of biogas plant with $H_2$ grid is obviously positive for production of bio- or e-fuels, due to both lower cost of electrolysis and simpler plant design and operation. The impact of RE sales on the production cost of the PtX products is positive as the RE sales contribute to cover part of the investment cost for the renewables. The production cost can be reduced up to 15% for high energy market prices and high RE sales with $H_2$ cost reaching as low as 68 €/MWh for high energy prices (2022) and a ratio between RE sales and energy for RFNBOs of one. These aspects are further discussed in figures 8 and 9.

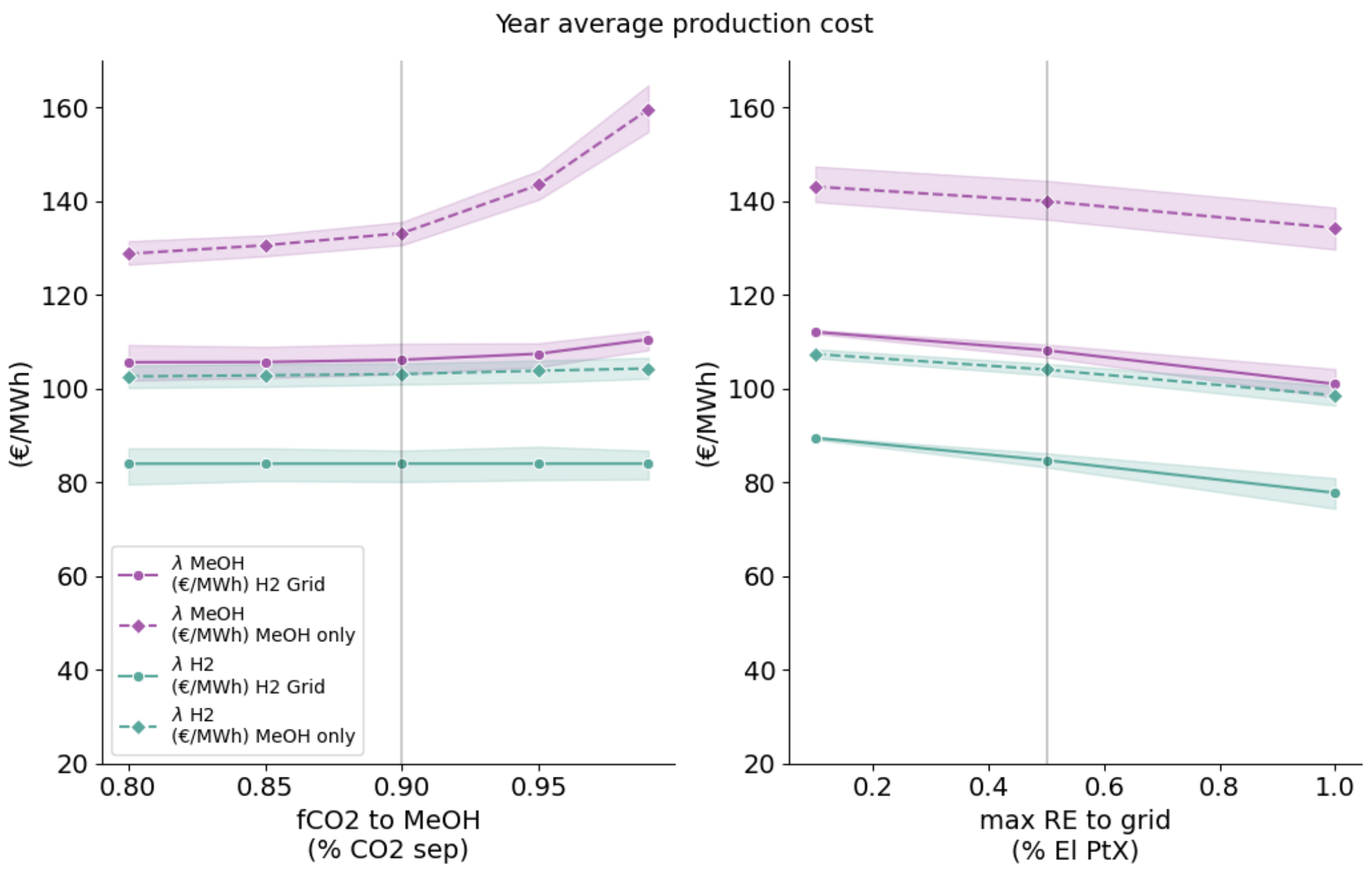

**Figure 4** – Production cost of MeOH and $H_2$ and sensitivity to ratio of $CO_2$ recovery and RE sales.

### 4.2 Impact of energy market prices and variability of prices in the behind the meter market

It is possible to reach similar production cost with in PtX hub optimized for low energy prices (2019) or high energy prices (2019), for both the $H_2$ grid and the MeOH standalone cases. The reason is that the application of RFNBOs legislation (post 2030) limiting the purchase of grid electricity to relatively rare conditions. Hence, the optimal system is based of additional renewables and becomes a net producer of RE to balance its operation. The internal market is therefore shielded against the price variations in the external grids as shows in figure 5, for the reference set of sensitivity parameters. The left side of figure 5 reports the annual distribution of shadow prices in the PtX park and the right side the variability in the external grid. The annual mean values are shown in the plots. This shielding effect also covers the internal heat market making which largely uses of the other sources than NG. Annual variability is also largely reduced compared to the external prices, which can set the basis for long term agreements between the companies in the park. A noticeable exception are those hours when the capacity of the electric grid connection (which is optimized) is saturated while the internal demand is high, leading to occasional spikes in the internal electricity price.

It should be noted that biomethane production cost is reduced (negative shadow price) as result of heat and electricity in the park being cheaper than the grid, making the integration profitable for the biogas plant without any changes in the operation.

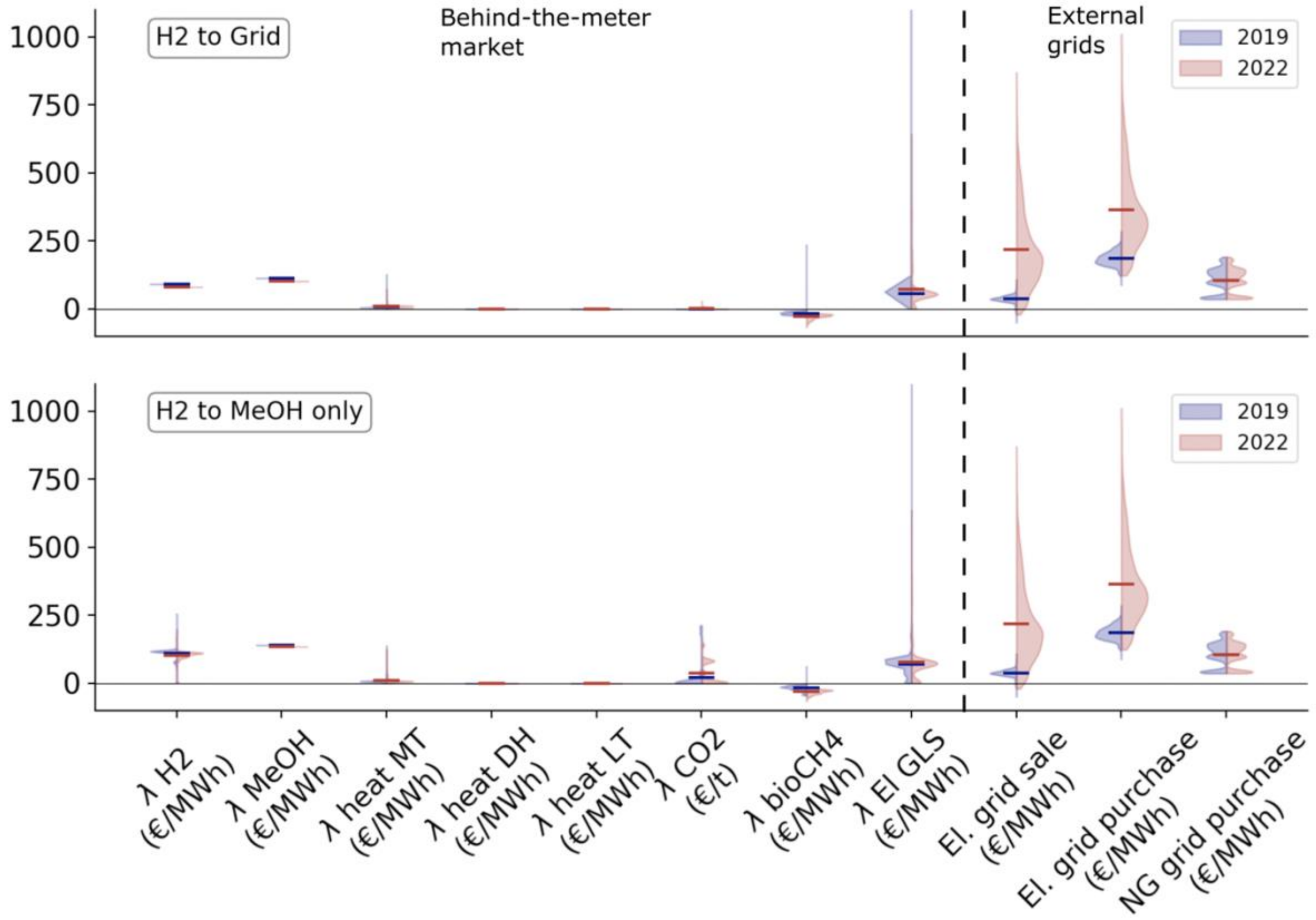

**Figure 5**: annual variability and mean shadow prices of commodities within GLS and the external grids. Energy years 2019 and 2022. Other parameters: rate of $CO_2$ to methanol: 0.9, maximum RE sales to grid: 0.5, $CO_2$ tax: 150 (€/t), DH and biochar production not enabled.

## 4.3 Total system cost

The optimal value of the objective function corresponds to the total annualized system (as defined in equation A1). A total system cost breakdown for the reference set of parameters is shown in figure 6 and for energy year 2022 and 2029. The total capital cost differs by a factor of 3 to 4 in the *H2 to grid* and *MeOH standalone* cases, with the renewables carrying most of the cost followed by the electrolysis plant, in comparison all the other areas require a minimal investment. Figure 6 also reports the total marginal costs for each case, which includes all costs and/or revenues for trading with external grids. It is worth reminding that if the optimization problem would not include sales at exogenously fixed prices, the total system cost will be fully recovered by the value of the loads ($H_2$ and MeOH) if traded at their shadow prices. However, in this formulation of the optimization problem, the revenues from the sales to the external grids (RE, DH and biochar credits), can lower the total system cost beyond the the value of PtX products. In figure 6 this is represented by the *total systems cost* which is the sum of capital and marginal costs and the *total systems cost with PtX sales* were the value of $H_2$ and MeOH was added. The results in Figure 6 are obtained for a *maxRE* of 0.5, which is sufficient in combination with 2022 prices to fully repay the capital costs of the PtX hub only via the sales of RE to the grid. This does not occur for a low energy price scenario (2019) where the sales of RE contribute only marginally to reduce the *total systems*. Overall, with 2019 energy prices both *$H_2$ to grid* and *MeOH standalone* plants are estimated to make a profit of about 4 M€/y when selling $H_2$ and MeOH at their production cost. High energy prices (2022) drive the installation of renewables to maximize the RE sales and the economics of the hub are heavily affected by RE sales, reaching a profit of 39 M€/y for the *$H_2$ to grid* case and 18 M€/y for the *MeOH standalone* case when selling $H_2$ and MeOH at their production cost.

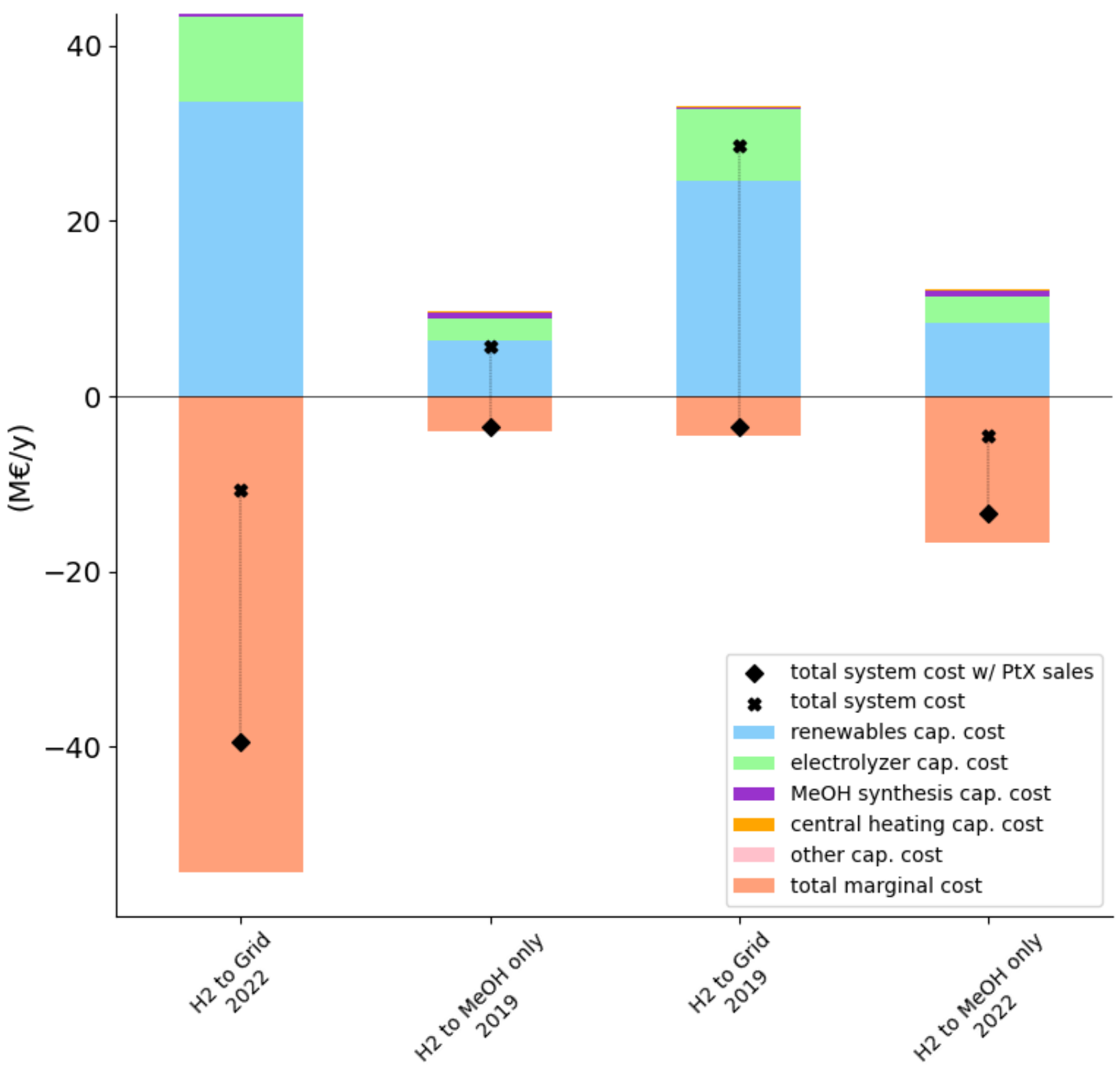

**Figure 6** – Breakdown of annualized total system. rate of $CO_2$ to methanol: 0.9, maximum RE sales to grid: 0.5, $CO_2$ tax: 150 (€/t), DH and biochar production not enabled.

## 4.4 Optimal system design – $H_2$ to grid

Although similar production costs can be achieved for 2019 and 2022 energy prices, the optimal design of the PtX hub can significantly differ. These variations are investigated in figure 7 (*$H_2$ to grid*) and 9 (*MeOH standalone*) which show the correlation between the year average shadow prices of the internal market in the upper triangle and correlations between the optimal capacities of the key plants and the annualized total system cost in the lower triangle. Figure 7 and 8 are based on the reference set of sensitivity parameters, and the results are grouped by the reference year for energy prices. The figures including the results of the full sensitivity analysis are available in appendix D.

In the *$H_2$ to grid* cases (Fig. 8) the optimal capacity of the renewables is driven by the revenues from the sales of RE to the grid. With low energy prices (2019) or limited sales of RE (*maxRE* 0.1) the PtX hub is powered by around 230 MW of wind power and 40 MW of solar which supply an electrolysis plant of about 125 $MW_{el}$. Under high energy prices (2022) the optimal capacity for renewables increases up to 380 MW of wind power and 150 MW of solar, with a higher capacity of electrolysis (up to 145 MW) to further increase the flexibility towards RE sales. Additionally, about 10-15 MWh of lithium-ion batteries provides flexibility for the internal electricity consumption. The capacity of the electrical connection with the external grid is linearly correlated to the RE export, varying from about 5 MW to nearly 100MW for the 2022 optimized case with high RE sales.

The optimal design of the production of MeOH process is simpler and tris operation smoother in all the *$H_2$ to grid* cases compared to the *MeOH standalone* ones (figures D5 and D6). Storages for $CO_2$ and $H_2$ are not required and the capacity of the MeOH plant is not affected by the external energy prices setting around 8 $MW_{meoh}$, indicating a smooth and continuous operation of the. This is the results of integrating a relatively small process (MeOH) with a large $H_2$ grid and recovering only 90% of the available $CO_2$. Regarding the heat network, the demand for MT and LT heat are mostly satisfied via the heat integration within the hub and only a small additional biomass boiler (0.8 $MW_{th}$ pellets) is installed to complement the generation, without the use of heat storage at any temperature level.
The shadow prices in the behind-the-meter market are affected by the external energy prices although the variations are considerably lower than those present in the external grids. Electricity and heat prices inside GLS are increased by higher external energy prices, due to the higher opportunity cost of selling that energy externally. Biomethane production cost is reduce considerably, especially for 2022 optimized solutions as grid purchasing of electricity and natural gas is avoided. However, low pressure $CO_2$ has nearly no value on the internal market unless it constrains the flexibility of the rest of the systems. This occurs only for very high energy prices and high RE sales, when RE export is competing with the operation of the PtX system.

## 4.5 Optimal system design – MeOH standalone

The optimal plant capacities and energy prices for the *MeOH standalone* cases are reported in figure 9. Optimal capacities for wind and solar for cases with low RE sales are about 50 MW and 25 MW of solar and increase up to 75 MW and 45 MW for 2022 prices and high RE sales. Battery capacity is very similar to the *$H_2$ to grid* case, indicating that it is used mostly to balance internal demand when energy prices are high. Compared to the *$H_2$ to grid* cases the methanol process required storage of $CO_2$ in the range of 200-350 tons, and $H_2$ for 70-110 MWh. The operation is more intermittent (figure D6) resulting in a higher capacity of the methanol plant for enhanced flexibility (12.5-15.5 $MW_{meoh}$), supported by 22 – 26.5 MW of electrolysis. Noticeably, the impact of RE sales on the optimal design is inverted between 2019 and 2022. For 2022 prices the sales of RE strongly contribute to reduce the total system cost, and the trade-off between smooth operation of PtX processes and high RE sales favors the latter, increasing the size of PtX components to add flexibility. For low energy prices the sales of RE are not relevant, hence the system is optimized to reduce investment cost in PtX with smaller capacities for MeOH production, electrolysis and storages, and smoother operation. Heat storage is not present for any temperature but, about additional 0.4 $MW_{th}$ of electric boiler and 0.8 $MW_{th}$ of pellets boiler are installed.
Energy prices in the internal market are affected by similar trends related to the external energy prices. For 2019 the prices of energy in the park are lower than 2022 and nearly unaffected by RE sales. On the other hand, RE sales have a relevant impact on the internal market with 2022 prices, but also reduce production cost of $H_2$ and MeOH similarly to the *$H_2$ to grid* cases.
It is worth mentioning that the MeOH standalone cases the low-pressure $CO_2$ separated at the biomethane upgrading fetches a relevant price (25-40 €/ton) as its scarcity would constraint the operation of the process resulting in even higher total cost.

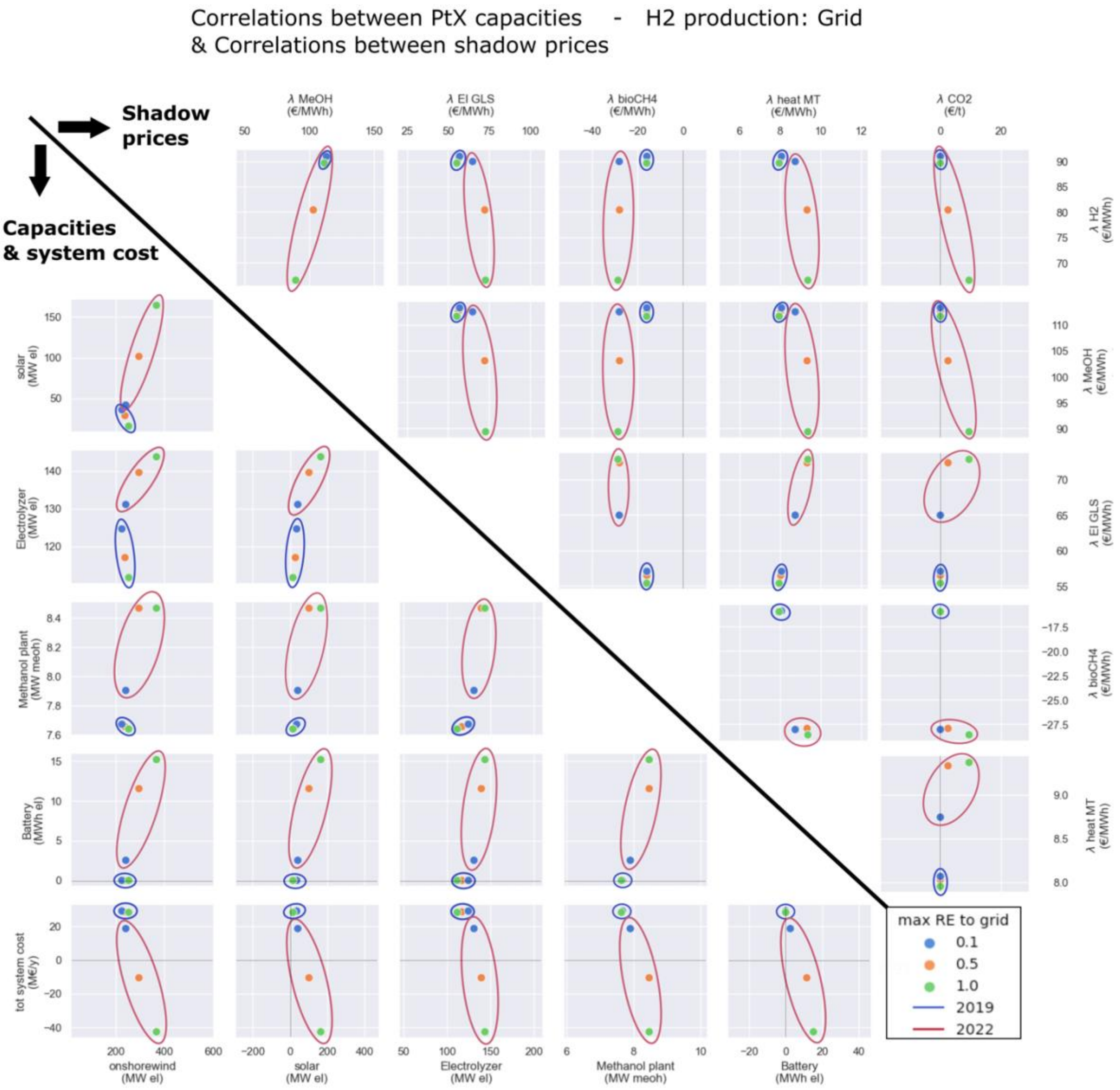

**Figure 7:** correlation between capacities in the optimal design of the PtX hub. Rate of $CO_2$ to methanol: 0.9, $CO_2$ tax: 150 (€/t), DH and biochar production not enabled.

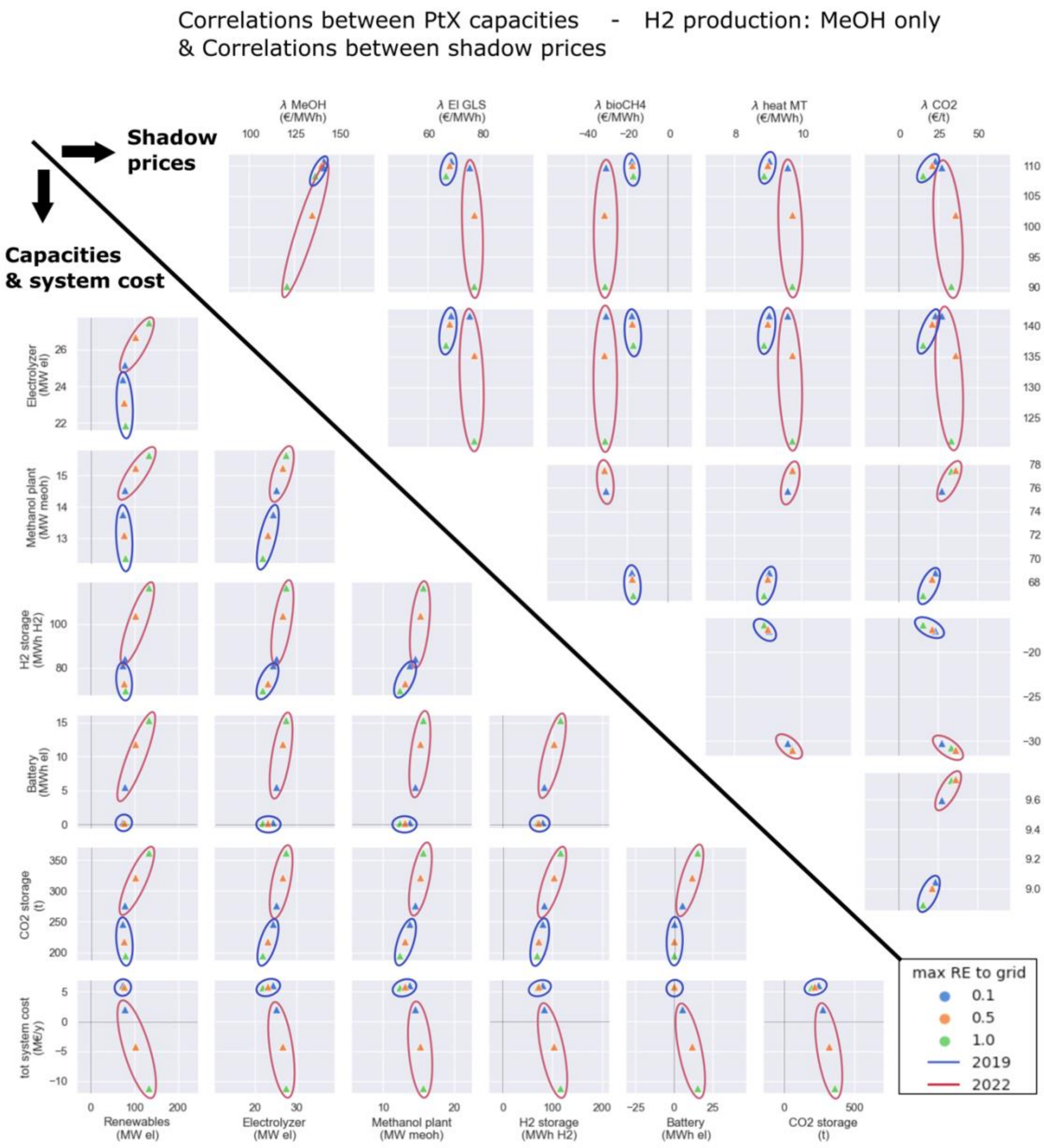

**Figure 8:** correlation between capacities in the optimal design of the PtX hub. Rate of $CO_2$ to methanol: 0.9, $CO_2$ tax: 150 (€/t), DH and biochar production not enabled.

## 4.6 Impact of DH and biochar credits

Enabling sales of DH or credits for biochar storage can increased revenues of the PtX and reduce total system cost. The impact on plant capacities and shadow prices in the internal market is reported in figures 9 and 10 respectively. The impact of DH sales and biochar credits is particularly dependent on the energy year because the DH price is set 54 €/MWh for both 2019 and 2022, and the biochar credits which are rewarded at the value of the $CO_2$ tax. Enabling DH generally corresponds to a further increase in renewables to power the heat pump system. Capacity of electrolysis if also increased as revenue from DH impact the production of $H_2$ slightly shifting it towards winter where DH can be sold (winter) due to higher demand. This effect is larger on the MeOH standalone cases where also the $H_2$ storage is increased to add flexibility for DH. The energy prices are affected, especially for the heat buses. Having an external market for heat increases the energy prices in the park by about 20 €/MWh, but reduces the slightly reduced the cost of MeOH and H2 in the *MeOH standalone* cases. Biochar is related to the GLS heat market as it is byproduct of heat generation form pyrolysis of the digestate fibers or straw pellets. Hence, the pyrolysis plant (SkyClean) plant substitutes the previous pellets boiler when biochar is rewarded with $CO_2$ credits. However, the revenue form biochar credits and the capacity of the pyrolysis plant are limited by the heat demand in the park, as in this model is not allowed to flare the pyrolysis gas without any use. The effect on biomethane is related solely to the avoided costs for NG in the existing boiler and grid electricity which increase with $CO_2$ tax.

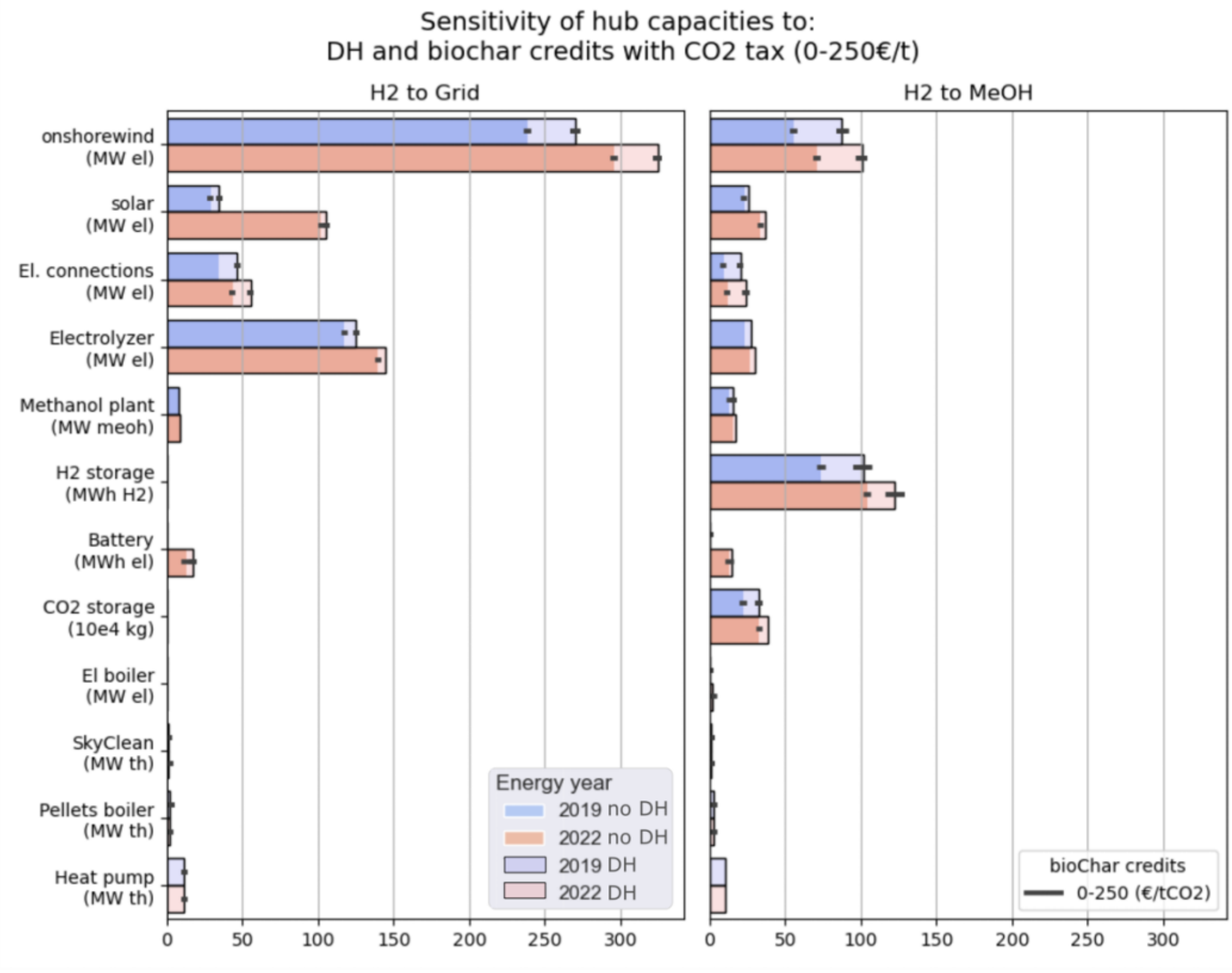

**Figure 9 -** Sensitivity of hub's capacities to DH and $CO_2$ tax with biochar credits

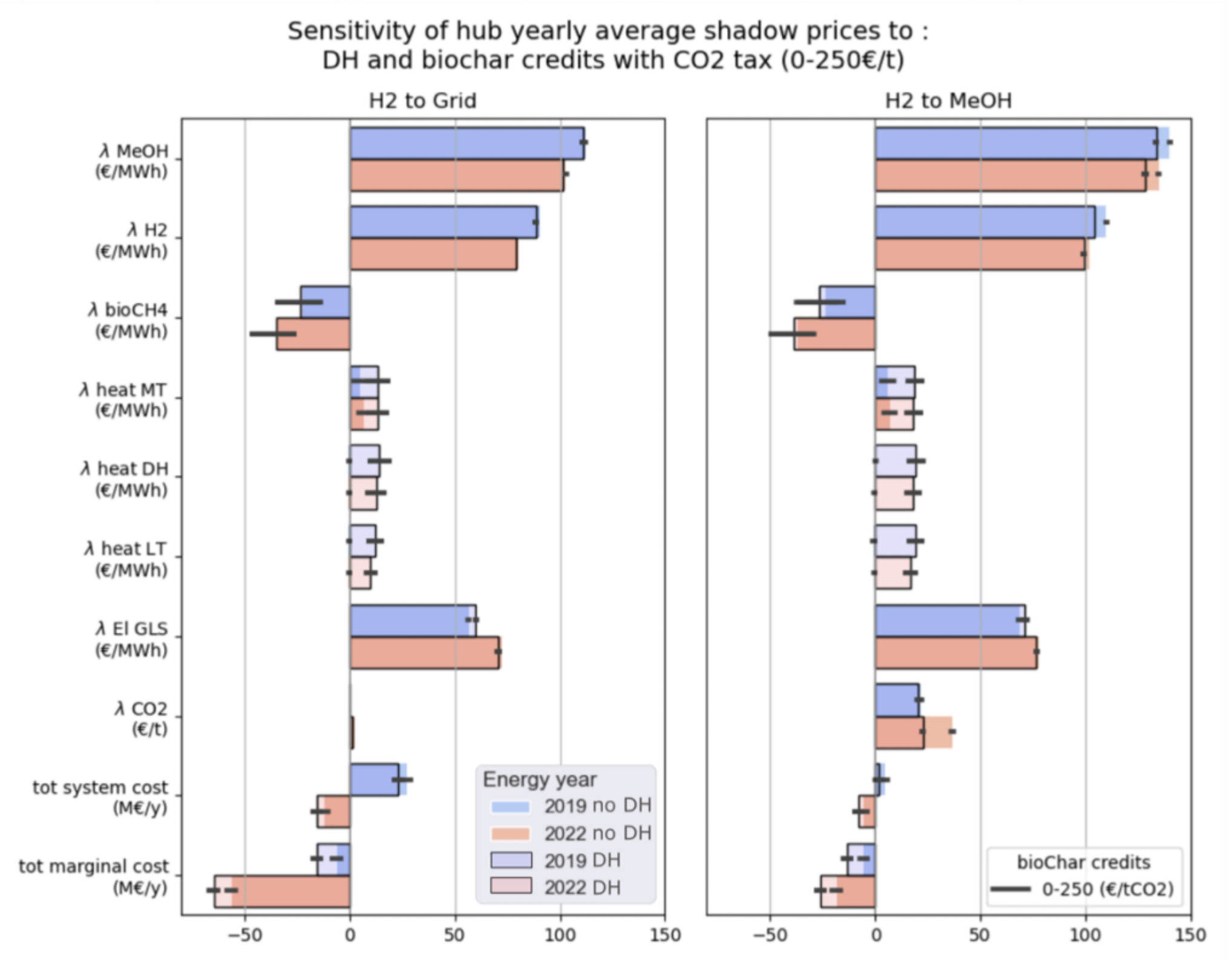

**Figure 10 -** Sensitivity of yearly average shadow prices to DH and $CO_2$ tax with biochar credits.

# 6 Conclusions

PtX hubs based on the integration of large electrolysis plant and biogas plants can provide a solid platform for exploiting largely unused biogenic $CO_2$ and satisfy the increasing demand for bio- and e-fuels at a competitive cost. The production cost for $H_2$ and MeOH in a PtX hub integrated with $H_2$ grid are estimated in 87-92 €/MWh and 111-114 €/MWh respectively. For comparison fossil methanol (priced at 360 €/t) would require a $CO_2$ tax of 180 (€/$t_{CO2}$) to reach a similar market price. Instead the optimized standalone PtX plant would produce MeOH and $H_2$ with 23% higher cost. The cost difference is largely due to the lower cost of electrolysis and to a significantly simplified PtX process without storage for $CO_2$ and $H_2$ and a smaller capacity of the MeOH plant and compressors. The ratio of $CO_2$ recovered to MeOH can have an impact on the production cost of MeOH especially for standalone plants, as exceeding 90% recovery would require further investments in storage.

The establishment of a behind-the-meter market as proposed in the GreenLab Skive hub, guarantees lower and more stable energy prices than the on external grids, independently on the external market prices. This is particularly valid with implementation of the RFNBOs regulations, prescribing additionality of renewables and limiting the use of grid electricity. The impact of sales of renewable electricity on the economics of the PtX hub depends largely on the energy prices of the external market. With energy prices as high as in 2022 the sales of RE can cover a substantial part of the capital cost of the PtX hub and further reduce production costs of $H_2$ and MeOH. The scope of this investigation did not concern any alternative use of the biogenic $CO_2$ than MeOH synthesis, hence their opportunity cost. Under this assumption the effect of $CO_2$ tax on the results is minimal since the system relies systematically on energy from the on-site renewables. However, if an alternative as $CO_2$ storage would be available, the value of the $CO_2$ credits is expected to influence the production cost of methanol.

# AKNOWLEDGEMENTS

This work was supported by research grant (40519) from VILLUM FONDEN.

# DATA AND CODE AVAILABILITY

The code to reproduce the results and visualizations is available on GitHub (https://github.com/BertoGBG/GLS_greenbubble), together with all input. Technology data assumptions were taken from github.com/pypsa/technology-data (v0.4.0). We also refer to the documentation of PyPSA (pypsa.readthedocs.io), for technical instructions on how to install and run the model.

# Appendix

## Appendix A – PyPSA model formulation

PyPSA is an open-source toolkit dedicated to simulating and optimizing modern power and energy systems. It encompasses a spectrum of functionalities including traditional generators and links, complete with unit commitment, as well as accommodating variable wind and solar generation, storage facilities, and coupling with other energy sectors. In this investigation only the *generator, load, bus, store* and *link* components were used for the formulation of the PtX hub model. *Storage unit* components were omitted instead a combination of *store* and *link* components was used. Connecting link are components which can convert energy carrier between buses and are used to model energy conversion processes (e.g chemical synthesis) using the multi-link formulation with multiple input and output.
Table 4 reports the mathematical formulation of the optimization problem.
In the formulation the index *s* indicates dispatchable generators, *t* indicated the time steps, *n* indicated buses (enforcing energy conservation), and *l* is the index for links. The optimization objective is the total annualized systems cost formulated as in Eq.1 in table A.1. Where $G_{n,s}$ is the power capacity of generators and $c_{n,s}$ are the associated fixed annualized costs. $E_{n,s}$ and $\hat{c}_{n,s}$ are the energy store capacity and its associated fixed annualized cost and $F_l$ and $c_l$ are the power capacity and associated annualized cost for connecting link *l*. The objective function is completed by the variable costs $o_{n,s,t}$ for generation $g_{n,s,t}$ and the variable cost $o_{l,t}$ for dispatch $f_{l,t}$ trough links, at every hour *t*.
The optimization is subject to a list of linear equality and inequality constraints. Eq.2 reports the constraint for energy conservation at each bus *n*, where hourly demand $d_{n,t}$ must be supplied by generators or imported from other buses via links. The energy flow of link ($f_{l,t}$) is multiplied by efficiency coefficient $\alpha_{n,t}$ indicating both the direction and the efficiency of the flow between the buses connected by the multilink.
The equality constraint expressed by Eq.2 are associated with the shadow price $\boldsymbol{\lambda}_{n,t}$, also known as Karush-Kuhn-Tucker (KKT), indicating the marginal price of the energy carrier in the bus.

The power dispatched by generators is constrained for every hour (Eq. 3) by the product of the between the installed capacity $G_{n,s}$ and the minimum and maximum availabilities $\underline{g}_{n,s,t}$ and $\overline{g}_{n,s,t}$. For renewables the minimum availability is zero and the maximum availability is the capacity factor at time *t*. The power capacity of all generators can be limited by the potential $\overline{G}_s$ related to physical and environmental constraints (Eq. 4), however in this formulation of the optimization problem the power capacity of generators and links were not constrained, except for the fixed capacity of the biomethane plant. Inequality constraints for links are described by Eqs 5 and 6. The maximum power flowing through the links is limited by their maximum physical capacity F. For bidirectional transmission links, $\underline{f}_{l,t}$ and $\overline{f}_{l,t}$ are equal respectively to -1 and 1. For energy conversion processes (unidirectional) $\underline{f}_{l,t}$ and $\overline{f}_{l,t}$ are equal respectivlely to 0 and 1. The power capacity of all links can be limited by the potential ($\overline{F}_l$) related to physical and environmental constraints (Eq.6), for example the charging and discharging of stores is controlled by links limiting the maximum power and setting the efficiencies ($\alpha_{n,l,t}$). The energy content of a store $e_{n,st,t}$ is constrained by the installed energy capacity $E_{n,st}$ (Eq. 7). The energy capacity of a store can be limited by other physical and environmental constraints (Eq. 8), e.g. the maximum allowed $H_2$ storage depends on the legal permitting at the site. Other additional technical constraints are the ramp-up and ramp down for generation and energy conversion technologies. These constraints are described in Eqs. 9 and 10 where the difference in generated or converted power between one snapshot and the other is limited by lower ($gdw_{s,t}$ and $fdw_{s,t}$) and higher ($gup_{s,t}$ and $fup_{s,t}$) bounds.

**Table A.**1 – Formulation of the general PyPSA-GLS optimization problem

| | |
|---|---|
| $\min_{G_{n,s},\ E_{n,s},\ F_l,\ g_{n,s,t},\ g_{n,l,t}} [\sum_{n,s} c_{n,s} \cdot G_{n,s} + \sum_{n,st} \hat{c}_{n,st} \cdot E_{n,st} + \sum_{l} c_l \cdot F_l + \sum_{n,s,t} o_{n,s,t} \cdot g_{n,s,t} + \sum_{l,t} o_{l,t} \cdot f_{l,t}]$ | a1 |
| **Subject to:** | |
| $\sum_{s} g_{n,s,t} + \sum_{l} \alpha_{n,l,t} \cdot f_{l,t} = d_{n,t} \leftrightarrow \lambda_{n,t} \quad \forall\, n, t$ | a2 |
| $\underline{g}_{n,s,t} \cdot G_{n,s} \leq g_{n,s,t} \leq \overline{g}_{n,s,t} \cdot G_{n,s} \quad \forall\ n, s, t$ | a3 |
| $0 \leq G_s \leq \overline{G}_s \leftrightarrow \mu_s \quad \forall\, s$ | a4 |
| $\underline{f}_{l,t} \cdot F_l \leq f_{l,t} \leq \overline{f}_{l,t} \cdot F_l \quad \forall\, l, t$ | a5 |
| $0 \leq F_l \leq \overline{F}_l \leftrightarrow \mu_l \quad \forall\, l$ | a6 |
| $0 \leq e_{n,st,t} \leq E_{n,st} \quad \forall\, n, st, t$ | a7 |
| $0 \leq E_{st} \leq \overline{E}_{st} \leftrightarrow \mu_{st} \quad \forall\, st$ | a8 |
| $-g_{dw,s} \leq g_{s,t} - g_{s,t-1} \leq g_{up,s} \quad \forall\, s, t$ | a9 |
| $-f_{dw,l} \leq f_{l,t} - f_{l,t-1} \leq f_{up,l} \quad \forall\, l, t$ | a10 |

# Appendix B – Technology assumptions in the model

## Alkaline Electrolysis

In alkaline electrolysis, electrodes commonly comprise materials such as steel, nickel, or nickel-plated steel, operating under pressures of up to 35 bars and temperatures ranging from 35 to 65°C. An integral feature is the utilization of a micro-porous diaphragm, effectively segregating electrode compartments to prevent gas mixing. The electrolyte employed is an aqueous solution of potassium hydroxide (KOH), with the electrode reactions summarized by equations b.1 and b.2. Crucial inputs for the electrolyzer are electricity and water, with water conductivity ideally maintained at approximately 1 μS/cm for Alkaline Electrolysis Cells (AECs). Consequently, an adaptable water purification system is imperative to accommodate variations in electrolyzer load. Typically, each electrolyzer unit is associated with a set of power electronics, varying in size and units. This set includes rectifiers and/or transformers, with rectifiers transforming AC current to DC current to deliver the appropriate DC current required for the different stacks. Water treatment requirements vary depending on the type of electrolysis and the water source. However, regardless of the specific case, a pre-treatment step followed by polishing within the water treatment process is essential. The choice of pre-treatment is dictated by the source of water, while the polishing step is determined by the electrolyzer technology. These steps may encompass filtration, aeration, UV treatment, desalination, and subsequent processes such as softening, demineralization, degassing, and electrodeionization (EDI).

$$2H_2O + 2e^- \rightarrow H_2 + 2OH^- \tag{b.1}$$

$$2OH^- \rightarrow \frac{1}{2}O + H_2O + 2e^- \tag{b.2}$$

**Table B.1**

| Input parameter | Value | Source |
|---|---|---|
| Pressure | 35 bar | [19] |
| Hydrogen Efficiency (LHV) | 62.2 % | [19] |
| Hydrogen production | 20.6 (kg/MWh$_e$) | [19] |
| Excess heat recoverable for DH | 22.3 % | [19] |
| Excess heat temperature | 50 °C | [19] |
| Water purity | 1 μS/cm | [19] |

## Biogas production and upgrading

The structure of the biomethane process is described here with aim to support and motivate the assumptions and used in this study, for a more general overview we refer to the Danish energy agency technology data catalogue for renewable fuels [41]. The process scheme for the biomethane plant is showed in figure B.2 in the configuration integrated with the SkyClean pyrolysis plant. Table B.2 reports the main mass and energy balance across the process and table B.3 reports other inputs concerning energy consumptions and conversion efficiencies. The biomethane production is divided in broadly two sections: biogas production where feedstock is converted to biogas and biomethane upgrading where methane is separated and cleaned for injection to the grid. The initial phase involves the reception and storage of biodegradable feedstock in pre-storage tanks, followed by subsequent processing in digesters reactors. These digesters are typically subjected to heating, either within the range of 35-40 °C for mesophilic digestion or 50-55 °C for thermophilic digestion. In the context of novel biogas plants integrated with gas upgrading, surplus heat from the upgrading facility serves as the energy source for the digesters. A prevalent choice among Danish plants is the utilization of continuous stirred-tank reactors, characterized by the continual extraction of a small quantity of digested biomass from the digesters, promptly replaced by an equivalent volume of fresh biomass, a procedure repeated multiple times daily. The processing time in the digesters, denoted as Hydraulic Retention Time (HRT), varies between 60 and 100 days, contingent on factors such as biomass input and technical specifications of the plants. The prevailing expectation for more recent plants is an HRT approximately around 65 days [24].

Following the biogas conversion process, the volume of the digestate remains relatively consistent, either maintaining parity or experiencing a slight reduction compared to the initial feedstock volume. The efficacy of biomass-to-methane conversion hinges on multiple variables, encompassing feedstock composition, processing duration, organic loading rate, and the precision of process control. Biomass components such as fatty biomasses, proteins, and specific carbohydrates (like sugars and starches) undergo relatively facile conversion into biogas. In contrast, only a fraction of cellulose undergoes conversion, with virtually no conversion observed for lignin. Approximately 46% of the dry matter in the feedstock transforms into biogas, with the remaining portion becoming part of the digestate, encompassing all organic matter not converted to gas. Post-separation into solids and fluids, this digestate proves recyclable as fertilizer. Subsequently, the gas undergoes treatment to align water and sulfur concentrations with desired levels, although this aspect is not illustrated in Figure B.2. Alongside the organic feedstock, the biogas process necessitates electricity for mechanical processing and heat for preheating and heating the reactor tanks reported in Table B.2.

Before introducing the gas into the gas grid, it is necessary to eliminate $CO_2$ content, thereby elevating its status to biomethane. The composition of raw biogas necessitates the removal of water moisture, particles, $H_2S$, $NH_3$, and $N_2$. Numerous technologies are deployed for carbon dioxide removal, including amine scrubbing, water scrubbing, membrane separation, and pressure swing adsorption. In this particular study, we adopted the assumption that the upgrading process relies on amine scrubbing. This technology employs amines that form chemical bonds with $CO_2$ and $H_2S$, effectively extracting them from the gas. The amine undergoes regeneration through the addition of heat, within the temperature range of 140-150 °C. Operating at pressures between 1-3 bar, the process necessitates a compressor to increase pressure for integration into the distribution gas grid (4/7 bar). The overall energy consumption for the upgrading section is considerably higher than for the biogas production as reported in Table B.2.

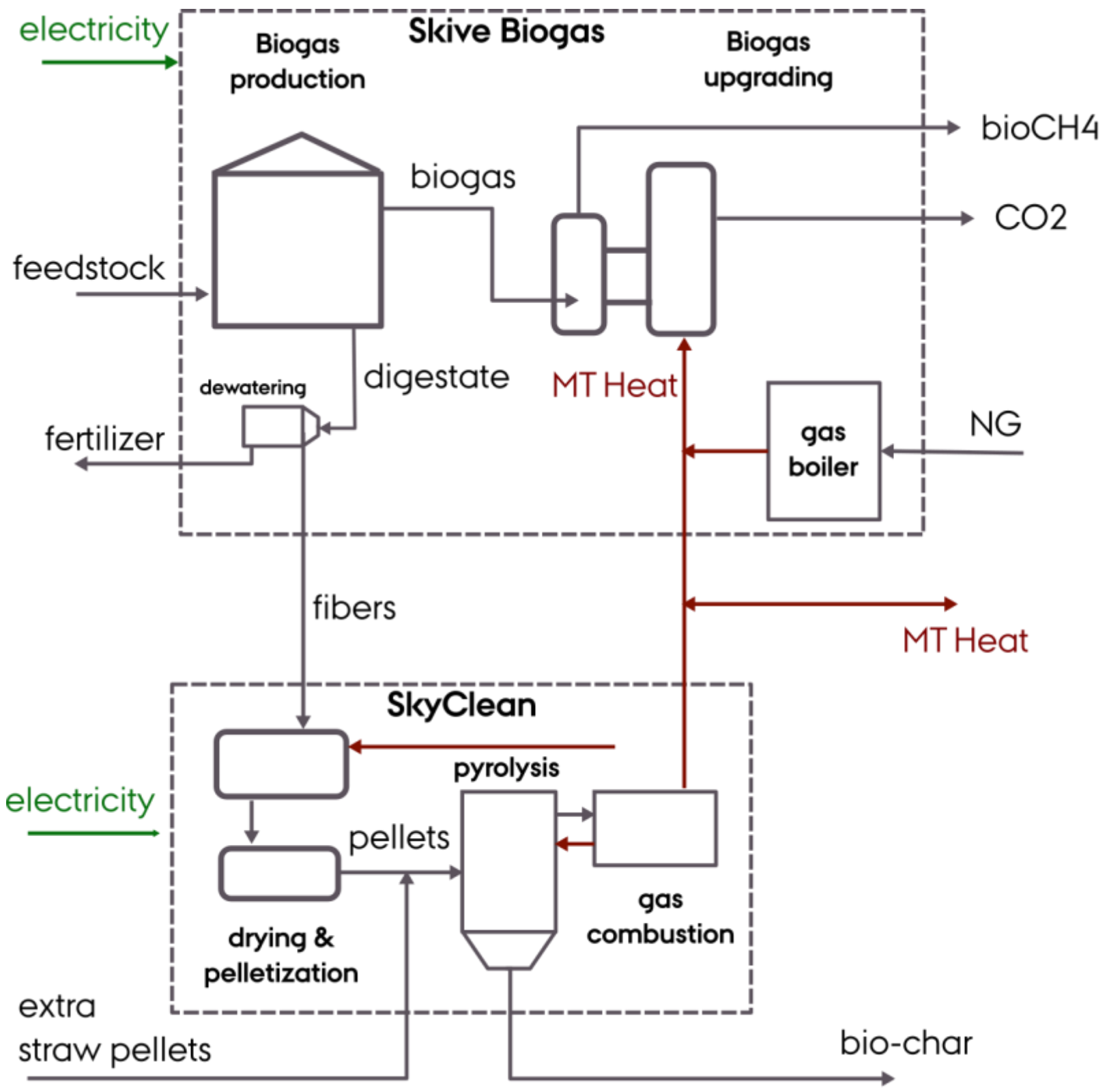

**Figure B.2** – Model schematics for the Biogas plant and SkyClean pyrolysis plant

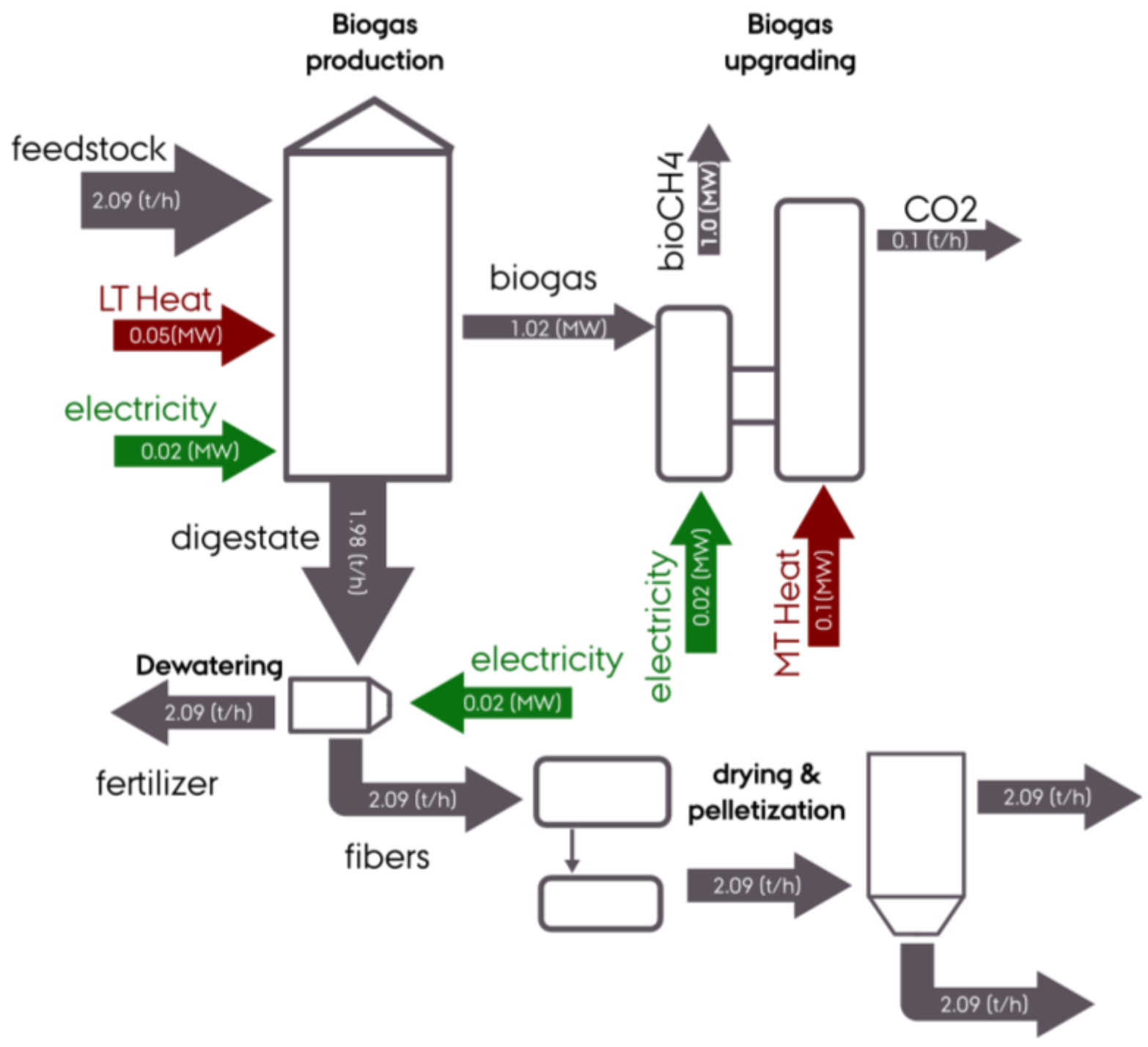

**Figure B.3** – mass and energy balance of biomethane plant including the pyrolysis of digestate fibers

**Table B.2** main assumptions about biomethane production

| Input parameter | Value | Source |
|---|---|---|
| **Biogas plant** | | |
| Biogas Skive biomass capacity | 500 000 (t/y) | |
| Biogas Skive biomethane production capacity | 240 GWh/y | |
| Feedstock type | 75 (%w) manure, (15%w) straw, (10%w) organic waste | Assumption based on [41] |
| Dry matter in feedstock | 12 (%w) | Assumption based on [41] |
| Conversion of dry matter in feedstock to biogas | 46 (%) | Calculated |
| Biogas composition | 65 (%v) $CH_4$, 35 (%v) $CO_2$ | Assumption based on [41] |
| Aux. process heat (50 °C) | 23.7 (kWh/$t_{input}$) | [41] |
| Aux. Electricity | 10.13 (kWh/$t_{input}$) | [41] |
| **Biomethane upgrading** | | |
| Technology | Amine scrubbing | |
| Heat demand (140°C) | 10.32% of biogas input | [41] |
| Electricity demand (inc. compressors) | 1.89 % of biogas input | [41] |

### Pyrolysis of digestate fibers

The SkyClean pyrolysis facility is designed to deliver medium-temperature heat to the Gas-Liquid-Solid (GLS) system while concurrently producing biochar for subsequent storage as a means of negative emissions. The primary feedstock for the pyrolysis process consists predominantly of digestate fibers, which undergo dewatering through a decanting centrifuge [59] [18], followed by drying and pelletization, or the utilization of straw pellets procured from the market (refer to Figure B.2). The process is grounded in slow pyrolysis within an updraft fixed bed reactor, facilitating thermochemical conversion of organic matter in an oxygen-deficit environment. While the pyrolysis process exhibits considerable flexibility regarding feedstock, meticulous pretreatment is required to ensure alignment with size distribution and moisture content requirements. Pre-treated biomass feedstock enters the pyrolysis reactor, undergoing heating up to temperatures within the 600 to 700°C range. The retention time in the reactor typically spans 5 to 10 minutes. Ideal feedstock dimensions consist of pellets ranging from 5 to 10 mm, with moisture content falling between 10 and 25%, contingent upon the specific feedstock employed. For the purpose of this investigation, it is assumed that the moisture content of both digestate pellets and straw pellets is 20%. Satisfying the heat demand of the pyrolysis reactor involves the combustion of a portion of the pyrolysis gases in an external reactor, ensuring the attainment and maintenance of the requisite pyrolysis temperature. The heat demand is variable and contingent upon factors such as feedstock moisture content. Literature [18] suggests an energy contribution of 9 to 13% from the incoming feedstock. Slow pyrolysis, as employed in this process, is deliberately tailored for achieving a high biochar yield characterized by substantial stability for carbon sequestration, specifically targeting a H/C molar ratio below 0.7 [43]. High-quality biochar as that produced at SkyClean have lower H/C ratio around 0.4 achieving larger C sequestration potential [43]. Numerous other parameters, detailed in [60], are essential for attaining optimal biochar stability and potential soil enrichment. The primary inputs and assumptions utilized in the model are documented in table B.3.

**Table B.3** main assumptions in pyrolysis process

| **Input parameter** | **Value** | **Source** |
|---|---|---|
| **Dewatering and drying** | | |
| Dewatering Technology | Decanting centrifuge | |
| Drying Technology | Rotary dryer | |
| Dry matter separation efficiency | 50.9 %w (fibers/digestate) | [41, 59, 61] |
| Electricity dewatering | 3.5 ($kWh/m^3_{digestate}$) | [41] [59] |
| Relative density digestate | 1.19 | [41] |
| Head demand for drying | 1 ($MWh_{th}/t_{H2O}$) | [59] |
| Dry mater in digestate | 6.8 %w | Calculated |
| Dry matter in fibers | 26 %w | [59] |
| Dry matter in pellets | 80 %w | [41, 43] |
| HHV dry fiber pellets | 17.6 (MJ/kg) | [43] |
| LHV dry fiber pellets | 16.0 (MJ/kg) | Calculated |
| **Ultimate analysis pellets from digestate fibers** (dry matter) | | |
| Ash | 15 %w | [41] |
| C | 43,6 %w | [43] |
| H/C molar | 1.5 | [43] |
| H | 5.4 %w | Calculated |
| N | 0.3 %w | [62] |
| P | 0.06 %w | [62] |
| O | 35.6 %w | Calculated |
| **Pyrolysis process** | | |
| Heat demand reactor | 13% of pellets energy (LHV) | [41] |
| Energy in Biochar | 38 % of pellets energy (LHV) | [41] |
| Energy in pyro Oil and Gases | 36 % of pellets energy (LHV) | [41] |
| Energy LT cooling | 9% of pellets energy (LHV) | [41] |
| Losses | 4% of pellets energy (LHV) | Assumption |
| Electricity process | 4% of pellets energy (LHV) | [41] |
| Carbon in biochar | 37% of pellets Carbon | [41] |
| H/C ratio in biochar | 0.4 | [43] |
| **Biochar sequestration** | | |
| Carbon sequestered (100 years) | 80 % of carbon in biochar | Assumption based on [60] |
| Price of straw pellets | 380 €/t | REF |

### Methanol process and compressors and $H_2$ and $CO_2$ storage

There are two main pathways for converting $CO_2$ to methanol [34] [16] [35] [36]. One pathway is to reduce $CO_2$ to carbon monoxide (CO) and use convectional catalyst for methanol synthesis, in a two-step process. A second and more efficient pathway is direct hydrogenation of $CO_2$ with hydrogen over a heterogeneous catalyst through a one-step process that converts $CO_2$ directly to liquid fuels. Methanol synthesis via catalytic hydrogenating of $CO_2$ with $H_2$ generally follows the set of reactions in equations B.3-5 and carried out at 50-100 bar and 200 - 300 °C. Two important competitive reaction are $CO_2$ hydrogenation (b.3) reverse water gas shift (b.4) complemented by CO conversion to methanol (b.5). The simplified process scheme for methanol synthesis via $CO_2$ hydrogenation is reported in figure B.2 and the energy and mass balance in table B.2. The generic description of the process includes pressurization of $H_2$ and $CO_2$, synthesis of MeOH and distillation of MeOH.

$$CO_2 + 3H_2 = CH_3OH + H_2O \quad \Delta H_0 = -49.16\ kJ/mol \tag{b.3}$$
$$CO_2 + H_2 = CO + H_2O \quad \Delta H_0 = 41.2\ kJ/mol \tag{b.4}$$
$$CO + H_2 = CH_3OH \quad \Delta H_0 = -91\ kJ/mol \tag{b.5}$$

Commercial process can be single- or multi step and include several recycling loops [17] [63]. The process design and operating condition of the methanol synthesis reactors (with pressure 50-100bar and temperature 200-200 C) are depended on techno-economic optimum for the specific process. The general argument is the limited yield of methanol are achievable when converting pure $CO_2$ even at elevated pressures compared to CO conversion [17]. Which leads to large recycle streams and consequently higher operating cost. Topsoe [64] and the Danish Energy Agency[41] indicate a pressure of above 80 bars for the process based on MK-317 SUSTAIN™ catalyst. The conversion efficiency is maximized when the reaction operates with an optimal stoichiometric ratio between $H_2$ and $CO_2$ equal to 3.

The inlet streams are compressed and pre-heated to before synthesis. The exit stream undergoes cooling to facilitate the condensation of the liquid product, primarily composed of methanol and water. This condensed mixture is then directed to a distillation column for the purification of methanol. The unreacted gases, predominantly hydrogen and carbon monoxide, are recirculated back to the reactor to enhance overall conversion efficiency. Subsequently, the methanol is temporarily stored in the liquid phase under atmospheric pressure. The mass and energy balance is based on the DEA data catalogue and adapted for decoupling the compressors from the rest of the process to allow for optimization of the capacity and operation of compressors and storages. The electricity demand for the process do not include the compressors for hydrogen and carbon dioxide but includes the compressor for recirculating and all other electrically driven components. DEA sets this consumption value to 100 $kWh/t_{MeOH}$ in line with other sources [52] [19] [21] [63] indicating that energy consumption for the recycling compressors is between 30 and 70 $kWh/t_{MeOH}$.

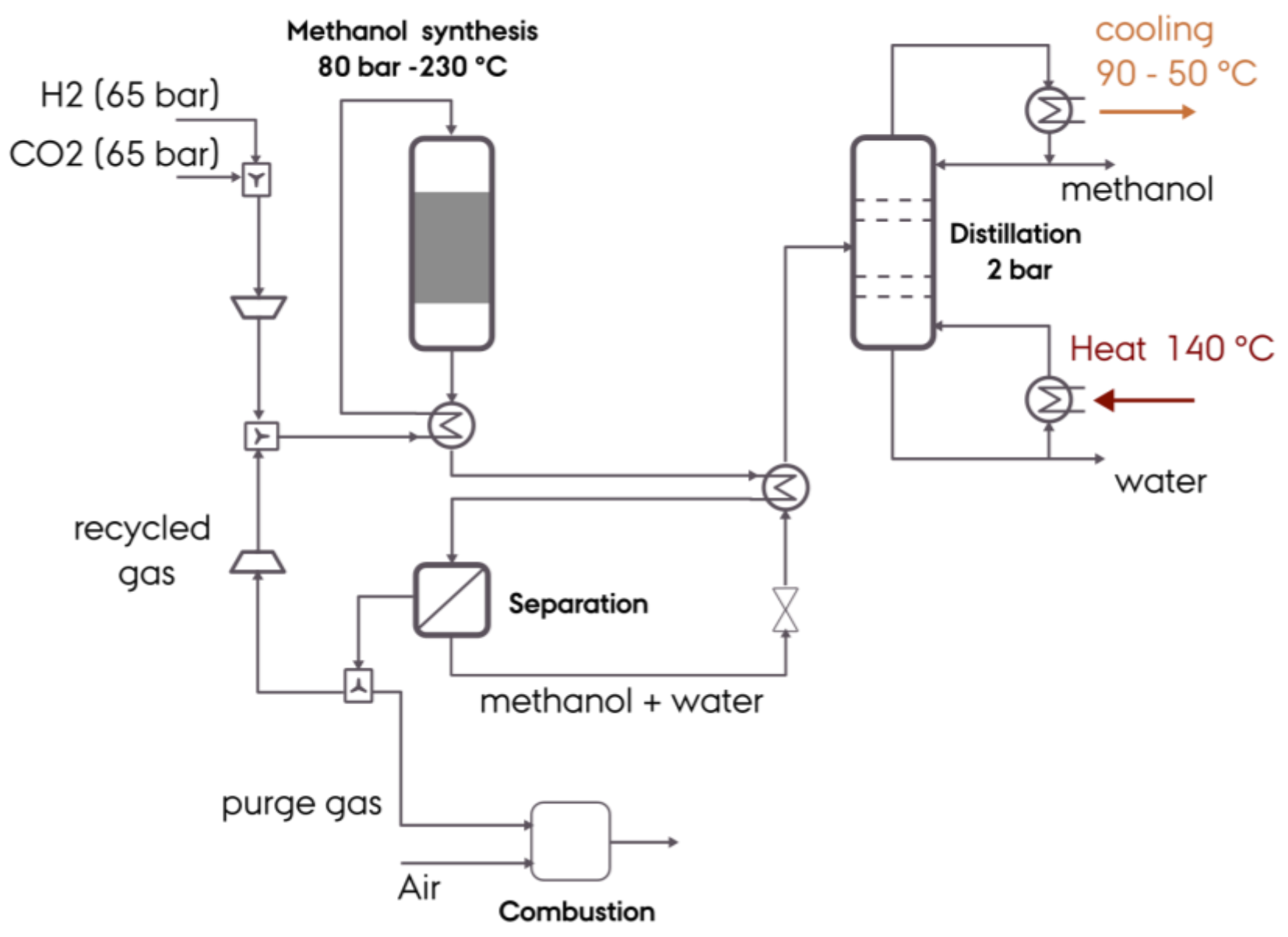

**Figure B.4** – Model schematics of the Methanol synthesis process

**Table B.4** – input parameters for MeOH synthesis

| Input parameter | Value | Source |
|---|---|---|
| MeOH synthesis pressure | 80 bar | [41, 64] |
| CO2 inlet flow | 1.4 (t/$t_{MeOH}$) | [41] |
| $H_2$ inlet flow | 0.19 (t/$t_{MeOH}$) | [41] |
| Heat demand (LP steam T < 140C) | 1.6 (t/$t_{MeOH}$) | [41, 64] |
| Heat excess (MP steam) | 0.67 (t/$t_{MeOH}$) | [41, 64] |
| Net heat demand (after heat integration) | 0.321 (MWh/$t_{MeOH}$) | [41, 64] |
| Heat to District heating (> 80 C) | 1.42 (MWh/$t_{MeOH}$) | [41, 64] |
| Electricity demand process* | 0.10 (MWh/$t_{MeOH}$) | [41, 64] |
| El. demand $CO_2$ compressor | 0. 096 (MWh/$t_{CO2}$) 1-80 bar | [19, 21, 51] [63] |
| El. demand $H_2$ compressor | 0.340 (MWh/$t_{H2}$) 35-80 bar | [52] [21] [45] |

* Excluding $CO_2$ and $H_2$ compressors but including recirculation to the MeOH reactor

The energy for compression of hydrogen from 35 to 80 bar was estimated to 340 kWh/$t_{H2}$ in line with [52] [21] for one stage reciprocating compressor and about 13% higher than the DEA estimation [45] for similar pressure ratio in large scale electrolysis plant for injection in hydrogen grids. Energy consumption for CO2 compression from 1 to 80 bar was estimated to 96 (kWh/$t_{CO2}$) for a four stage intercooled compressor in line with [51] [21] [63].

Hydrogen can be stored in a tank with maximum pressure of 70 bar, the system is a standard commercial hydrogen tank for balance of operation in electrolyzer plants. The typical capacity for such tanks is up 500 kg (ref dea). The charging and discharging rates assumed to be higher than the compressor output rate and the MeOH plant input rates controlled by the final compression of $H_2$ and $CO_2$ mixture to 80 bars. Hence charging and discharging rate are not assumed to constraint the system. Additional compression work to store hydrogen is required compared to direct compression, which is estimated to 0.068 (MWh/$t_{H2}$). Stand-by losses due to permeation are neglected.

The $CO_2$ storage in a system of high-pressure cylinders or high-pressure tank has a maximum pressure of 60 bars at room temperature. The total capacity of the systems is not constrained in the model. The maximum charging and discharging rate of the system are depended on the design of the manifold and flow rate of the compressors. It is here assumed that charging and discharging rates do not constraint the system as it can be designed to match the flow rates of the compressors. However, additional compression work it is required to charge and discharge the storge which is estimated to 10 kWh/$t_{CO2}$.
$CO_2$ liquefaction and storage in tanks is a convectional for biogas plants shipping $CO_2$ via truck. Prior to liquefaction impurities in the as hydrogen sulfide and volatile organic compounds must be removed and moisture is removed during the compression process. The system is composed by a liquefaction section where $CO_2$ is compressed then cooled to – 27 °C via an ammonia refrigeration cycle and a storage section with an insulated tank at 16 bars. The pressure in each tank is managed by controlling the $CO_2$ to boil-off the rate depending upon the surrounding temperature.

**Table B.5 -** inputs to models of hydrogen and carbon dioxide storage

| | | |
|---|---|---|
| **Hydrogen storage** | | |
| pressure | < 70 bar | [47] |
| temperature | < 50 °C | [47] |
| Tank typical capacity | 500 kg | [47] |
| Additional compression energy | 0.068 ($MWh/t_{H2}$) roundtrip | |
| **$CO_2$ liquefaction & storage** | | |
| Liquefaction Capacity (typical) | 2 (t/h) | [45] |
| Storage pressure | 16 bara | [45] |
| Evaporator pressure | 16 bara | [45] |
| Evaporator annualized cost | 3.76 k€/y/ ($t/h_{CO2}$) | internal |
| Storage temperature | -27 °C | [45] |
| Electricity for liquefaction | 61 ($kWh/t_{CO2}$) | [45] |
| Heat out of refrigeration | 166 ($kWh/t_{CO2}$) | [45] |
| Temperature of heat out of refrigeration | 80 °C | [45] |
| Typical Capacity for bigas plants | 1 – 4 (t/h) | [45] |
| **$CO_2$ storage in cylinders** | | |
| Pressure | < 65 bar | |
| Temperature | < 50 °C | |
| Additional compression energy | 10 ($kWh/t_{CO2}$) roundtrip | |

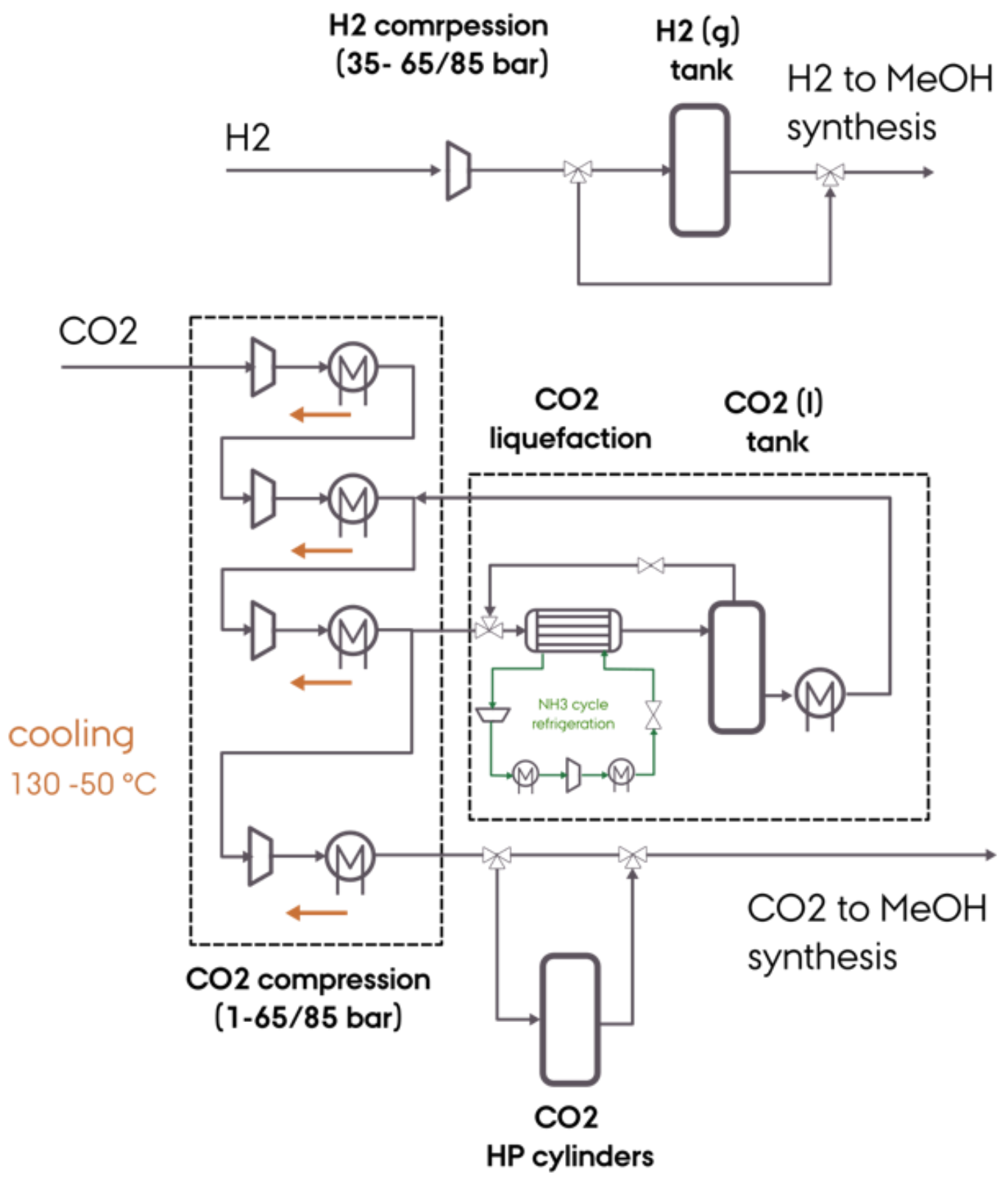

**Figure B.5** – Model Schematics of the $CO_2$ and $H_2$ compression and storage

The boiled-off $CO_2$ can be sent back for re-liquefaction. The liquefaction section is based on closed loop refrigeration cycle using ammonia as working fluid. The process is integrated with the main $CO_2$ compressor (Fig B.3) and the main energy consumption for the liquefaction is driven by the refrigeration loop (REF DEA). At the discharge of the storage the carbon dioxide is pumped and evaporated at 16 bar to reach the conditions for mixing with hydrogen in the MeOH process, the cost for the evaporator is added to the that of the liquefaction plant from DEA as not included in the reference. It must be considered that in absence of a low temperature cold utility on site is not possible to recover the cold available when the re-heating the $CO_2$ from the storage conditions. The DEA considers economically feasible for liquefaction plants to have capacity between 1 t/h and 4 t/h, and commercial offers are available up to 15 t/h, which was used as a constraint for charging and rate in the model (tab).

**Symbiosis net and Energy Storage Technologies**

The term symbiosis net refers to the infrastructure enabling behind-the-meter trading of electricity, heat, biomethane, carbon dioxide and hydrogen. It includes the additional piping, connection points, and energy storage technologies for electricity and heat instead storage of carbon dioxide and hydrogen is allocated to the plants using them (e.g. methanol plant). The simplified schematics of the electric grid at GLS is reported in Figure B.6. The figure reports the possible point of connection and internal transformers in the superstructure of the model. Two main electrical buses are available at GLS one at 20kV for powering the variable loads as the electrolysis plant and another one at 10kV for the other steady industrial loads. Both buses are connected to the grid and the renewables (on the 10kV bus). Electrical energy storage is possible though lithium-ion battery placed on the 10kV bus. The maximum C-rate for charging and discharging of the battery is limited to 1 due to time resolution in the optimization. The average heat loss in the network was assumed to be 3% and it is allocated to the heat exchangers. The technical inputs and cost for industrial heat pumps refer to the DEA technology catalogue. The connection to the external district heating network is not included in the GLS network and it is assumed to be provided by the local DH operator. The cost for piping of hydrogen and $CO_2$ was obtained from the DAE catalogue. Both flows are assumed of being gas phase and relatively low pressure (35 bars for hydrogen and 1-5 bars for $CO_2$).

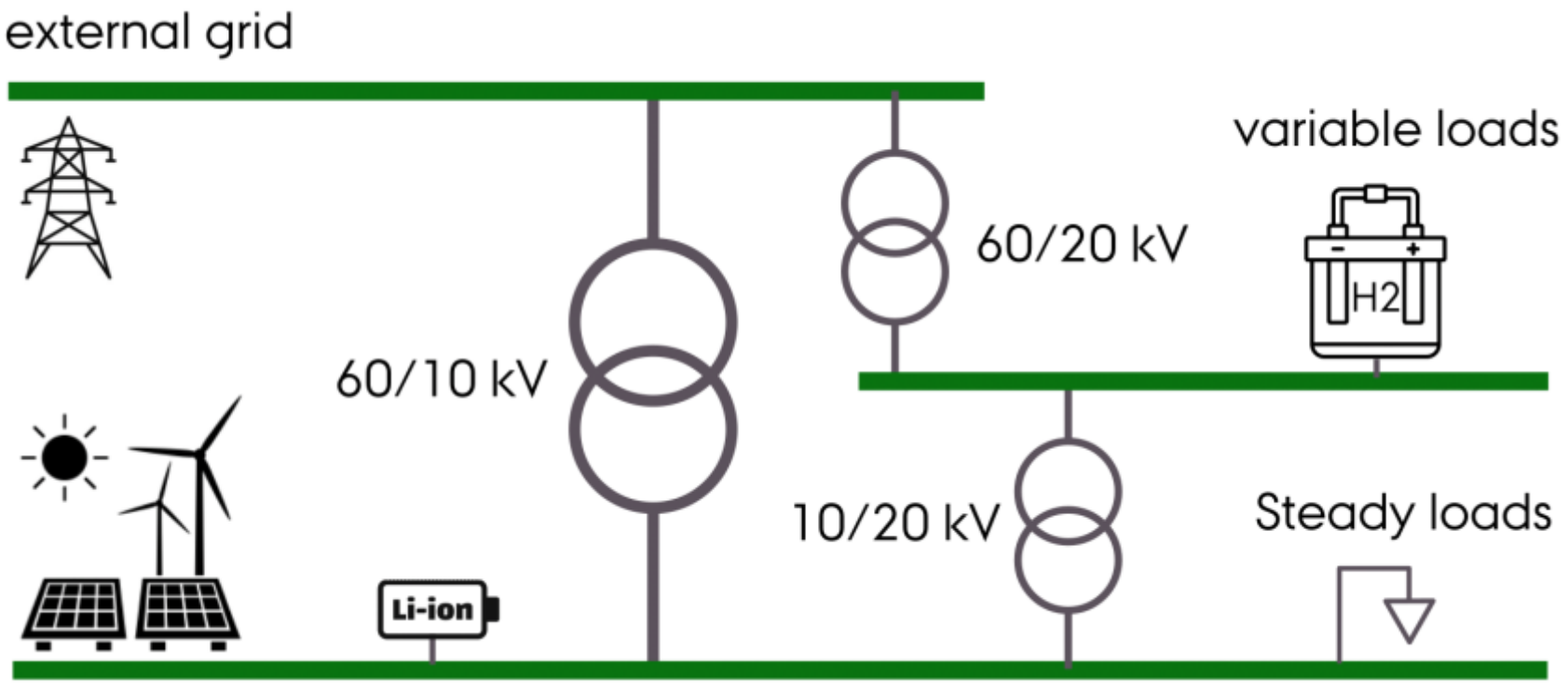

**Figure B.6** All available transformers configuration in GLS model.

**Table B.7**

| Input parameter | Value | Source |
|---|---|---|
| Li-ion battery maximum C-rate | 1 | assumption |
| Li-ion battery charging efficiency | 96% | [19] |
| Li-ion battery discharging efficiency | 96% | [19] |
| Thermal battery round trip efficiency | 95% | [49] [53] [54] |
| Hot water tank round trip efficiency | 98% | [19] |
| Assumed length of heat networks | 5 km | assumption |
| Heat pump average COP | 2.7 ($MW_{th}/MW_{el}$) | [44] |
| Assumed length of $H_2$ network | 1.5 km | assumption |
| Assumed length of $CO_2$ network | 1.5 km | assumption |
| Efficiency of biomass boiler | 91% (LHV) | [44] |
| Efficiency of NG boiler | 98% (LHV) | [44] |
| Efficiency of Electric boiler | 99% | [44] |

* applied to the PYPSA link component representing heat exchangers

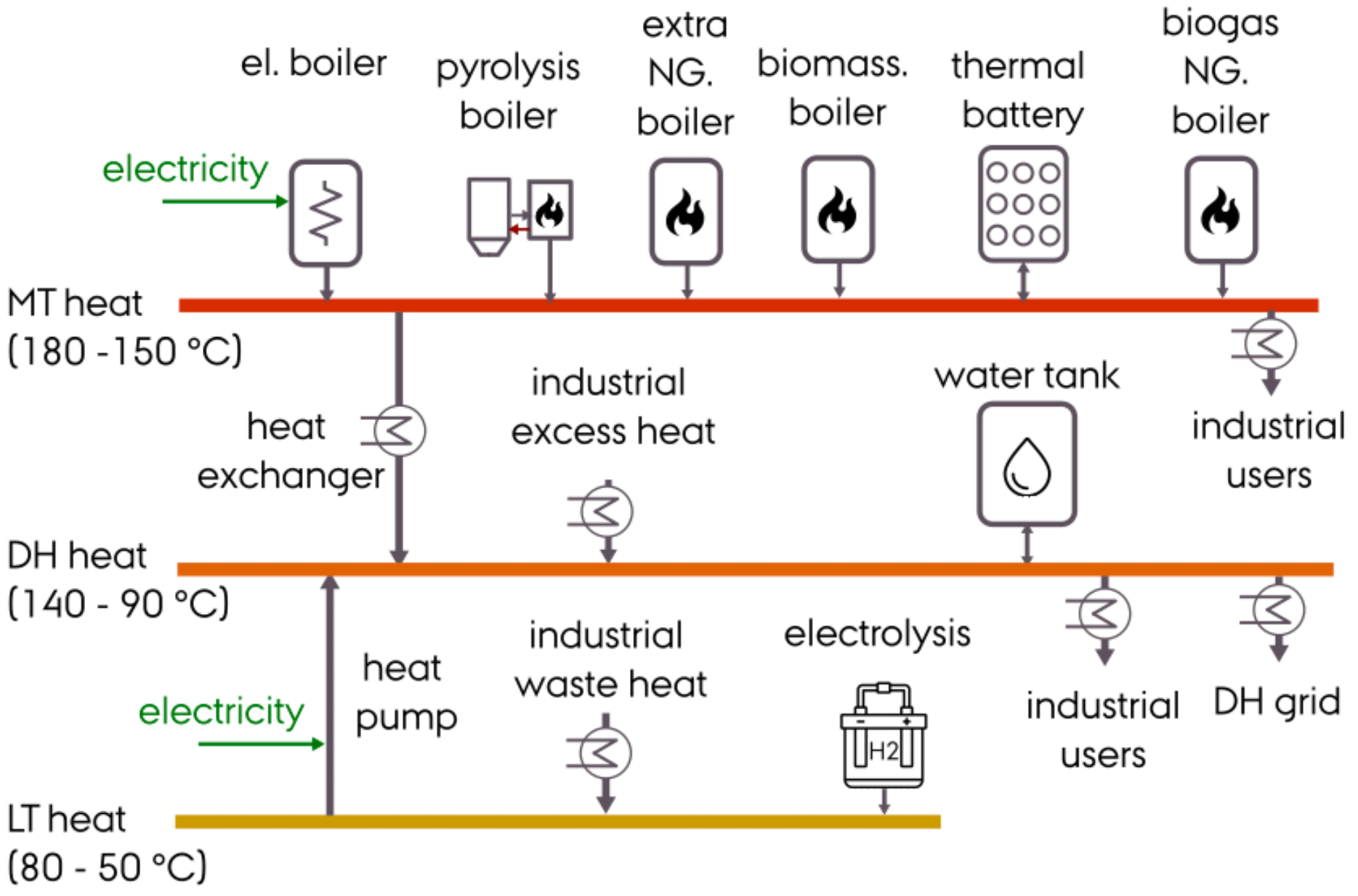

**Figure B.7** – Model schematics of all possible connections in the heat network

## Appendix C - external market prices

This section reports the time series and load duration curves for input data at the interface between GLS and the large energy systems: Electricity prices (spot prices, spotprices including tariffs for TSO [57] and DSO [58] and $CO_2$ tax), NG price including $CO_2$ tax, demand for DH, demand for electricity in DK1 (only the profile of the time series is used as input).

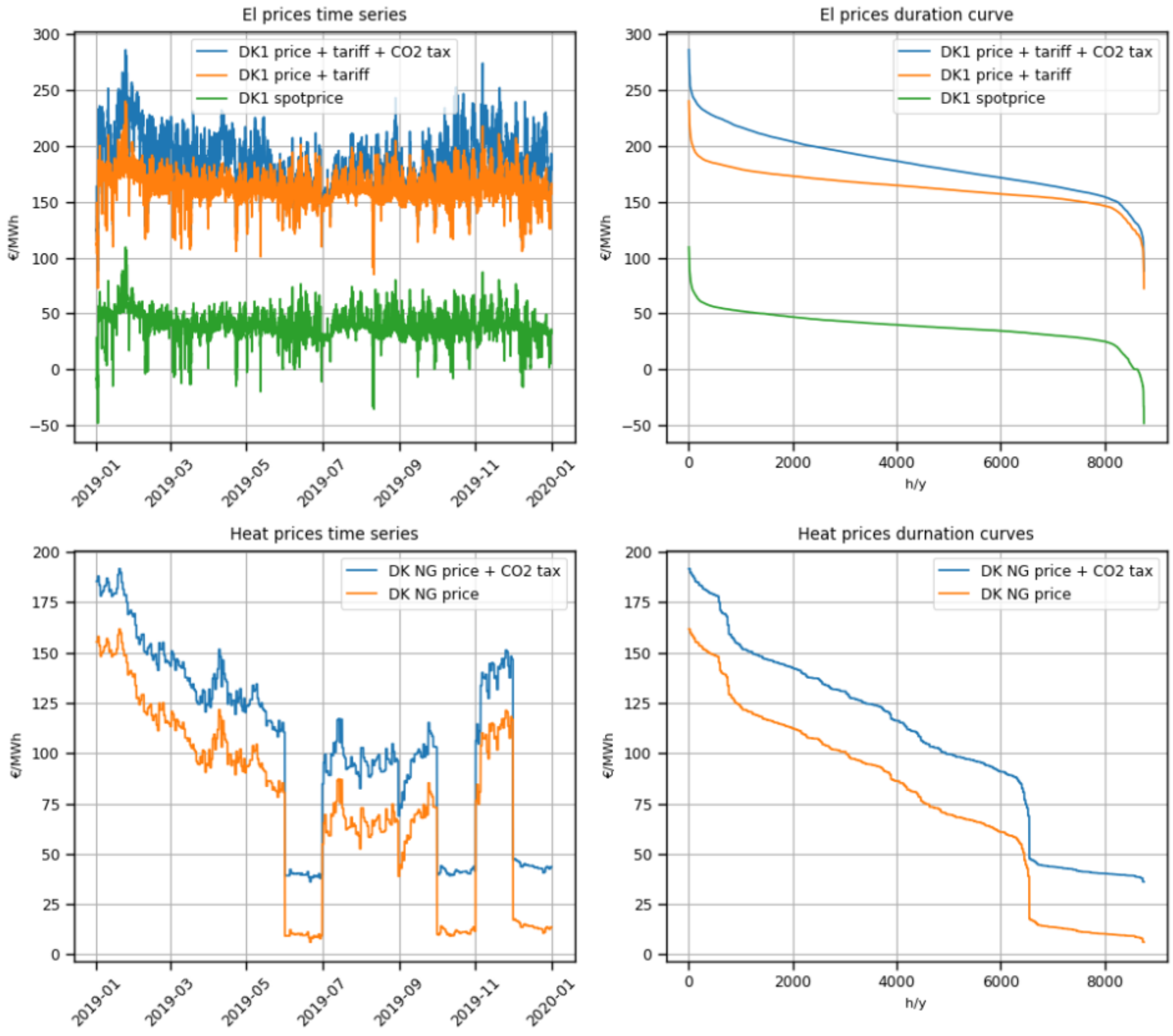

**Figure C.1** – Energy year 2019: Duration curves for electricity and natural gas prices [56]

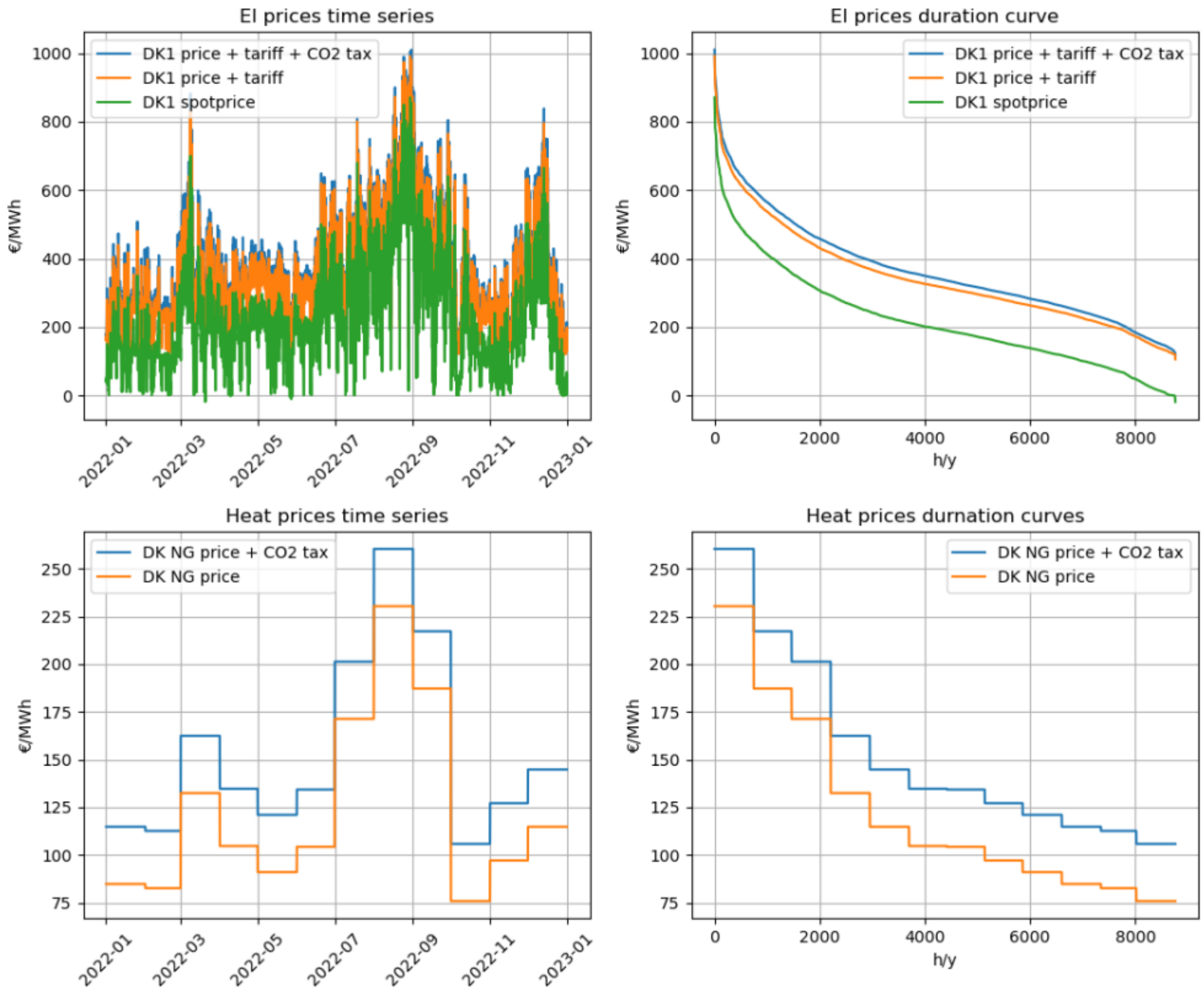

**Figure C.2** – Energy year 2022: Duration curves for electricity and natural gas prices [56]. NG prices have different time resolution than 20019 due to a change at the data source.

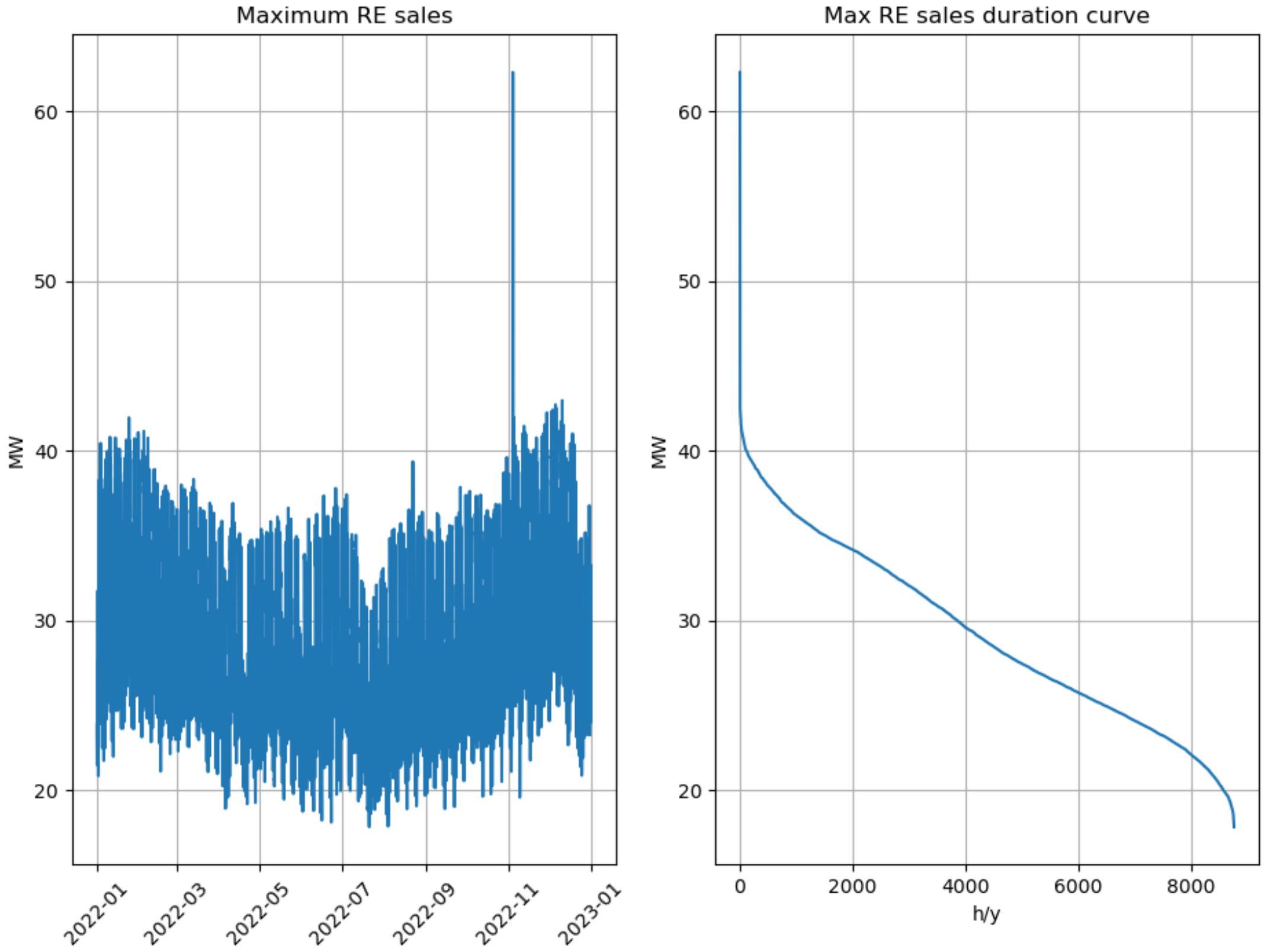

**Figure C.3** – Example of time series and duration curve for maximum sales for RE (external electricity demand) for *maxRE* equal to 0.5. Based on electricity demand in DK1 market [65].

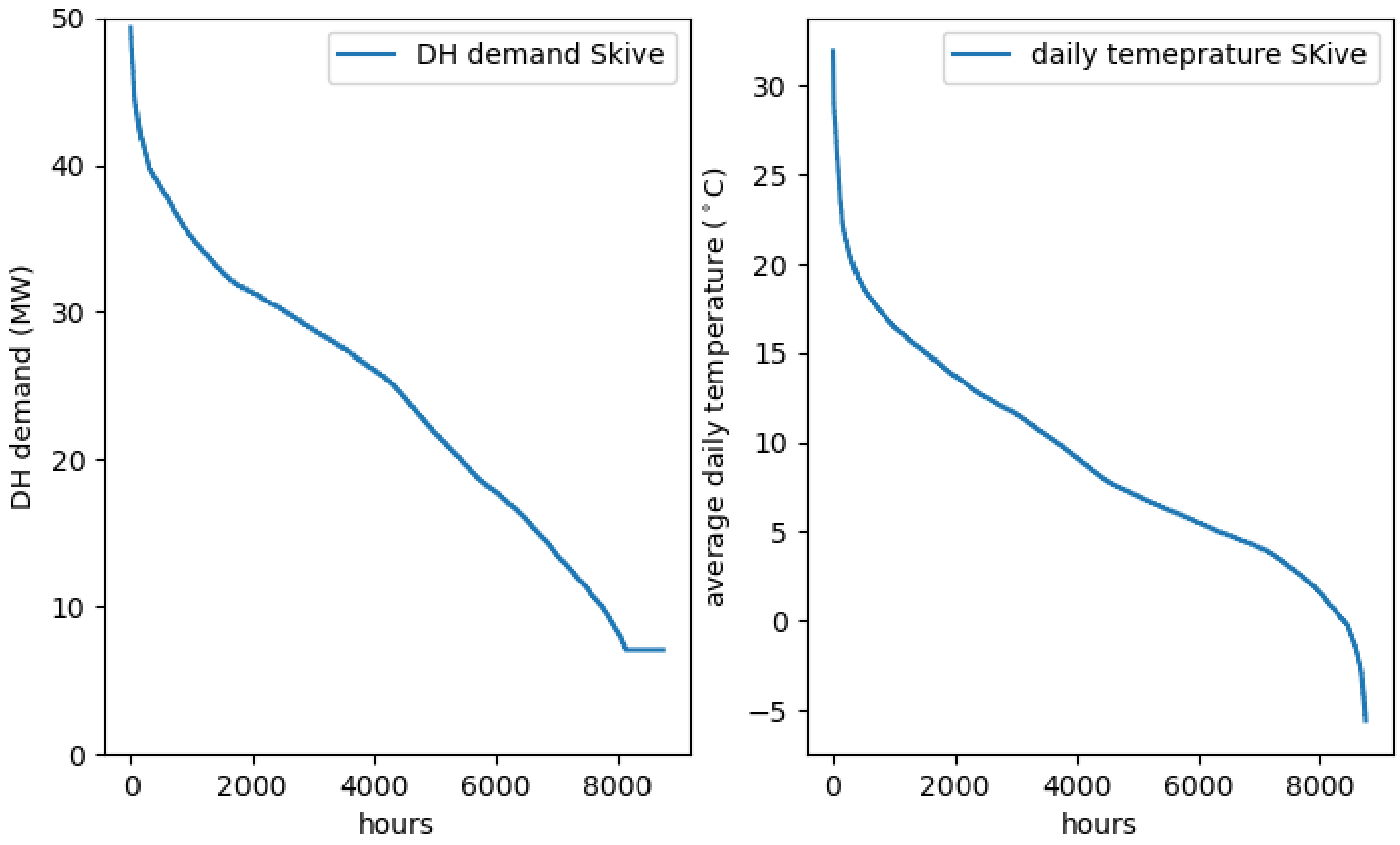

**Figure C.4** - Duration curve for District heating demand and average daily temperature in Skive (DK)

## Appendix D – Results: correlations between hub's capacities and between shadow prices

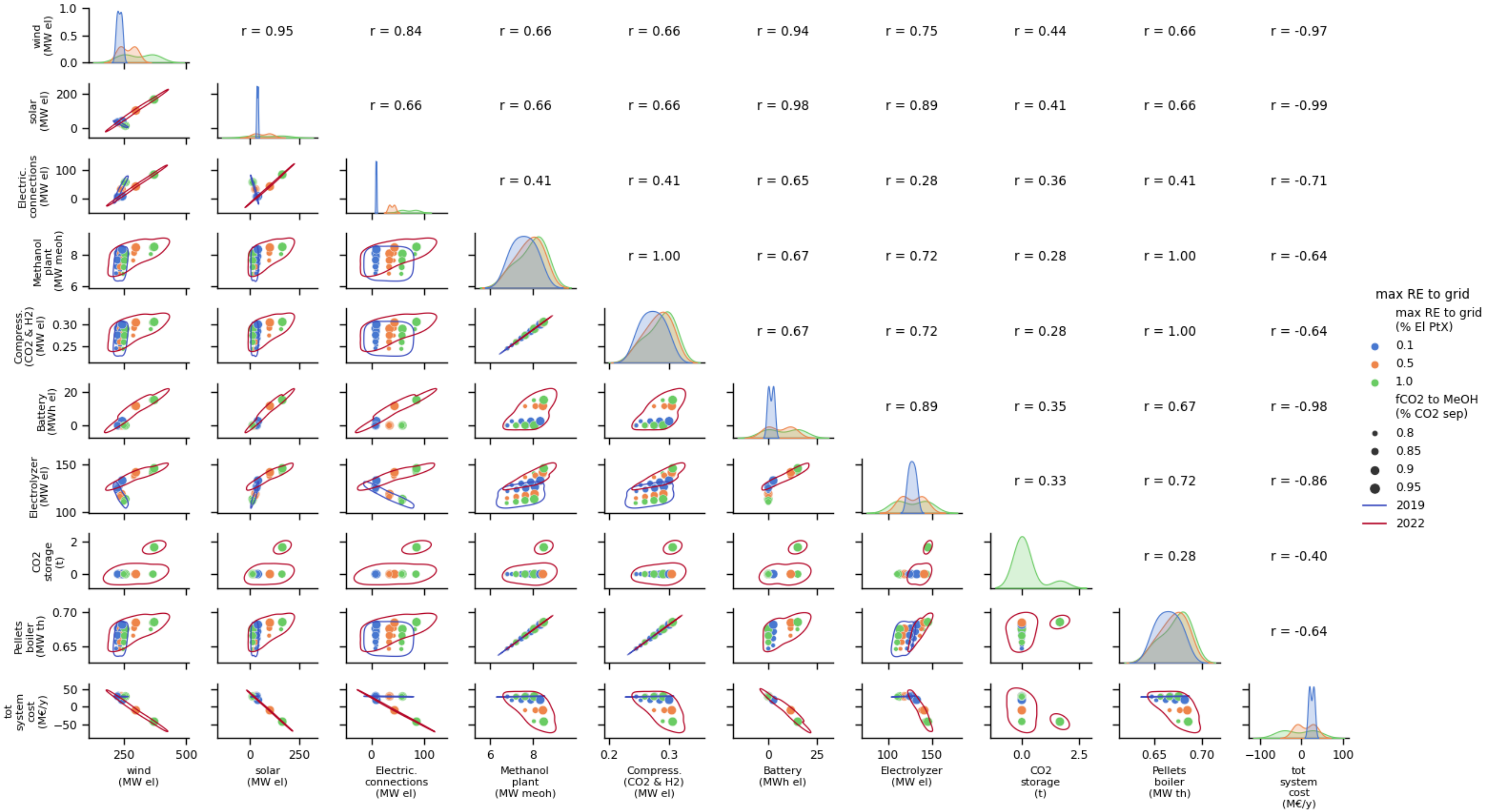

**Figure D1** – *$H_2$ to Grid* scenario: correlation between capacities in the optimal design of the PtX hub. $CO_2$ tax: 150 (€/t), DH and biochar production not enabled.

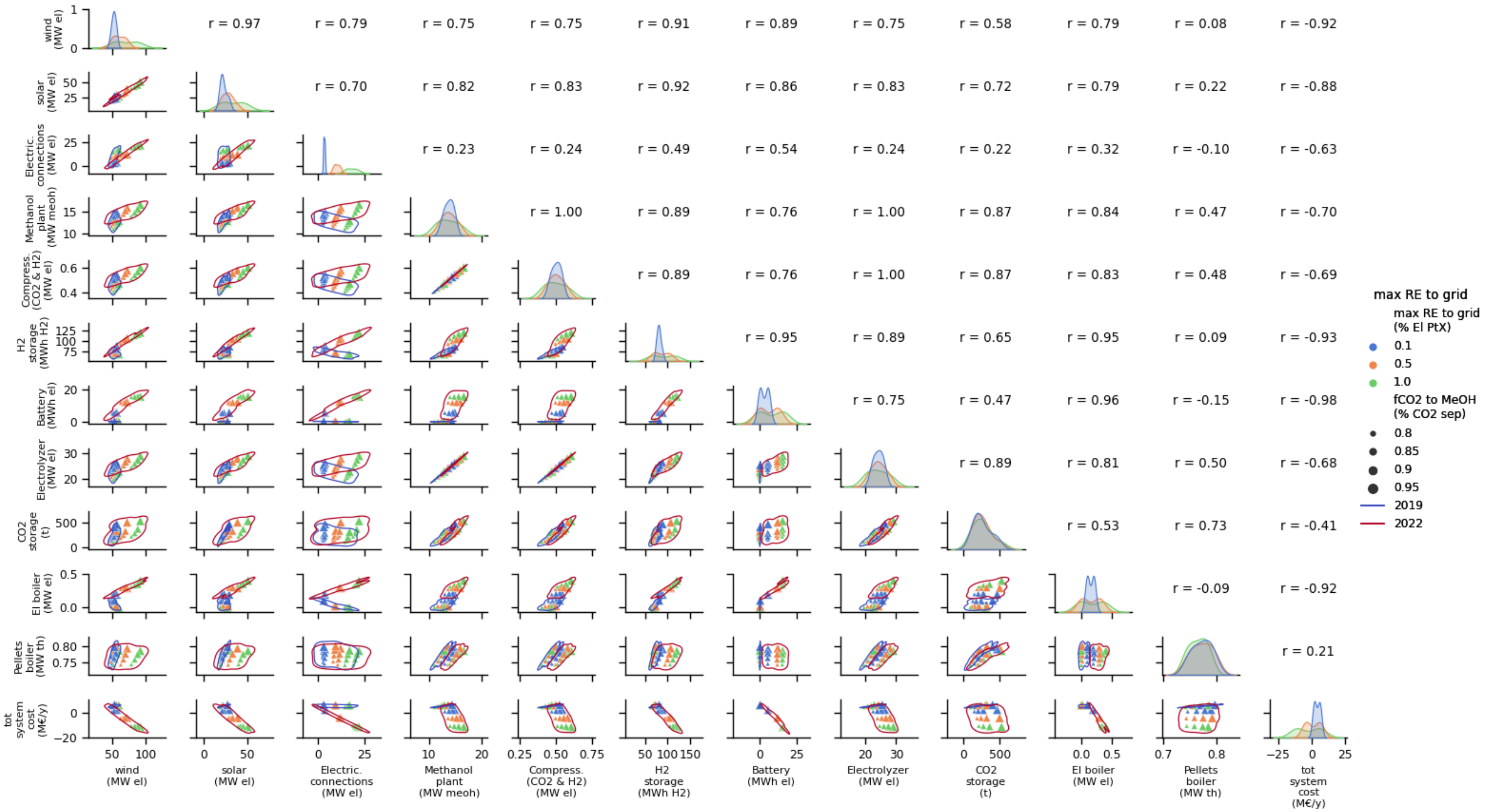

**Figure D2** – *MeOH standalone* scenario**:** correlation between capacities in the optimal design of the PtX hub. $CO_2$ tax: 150 (€/t), DH and biochar production not enabled.

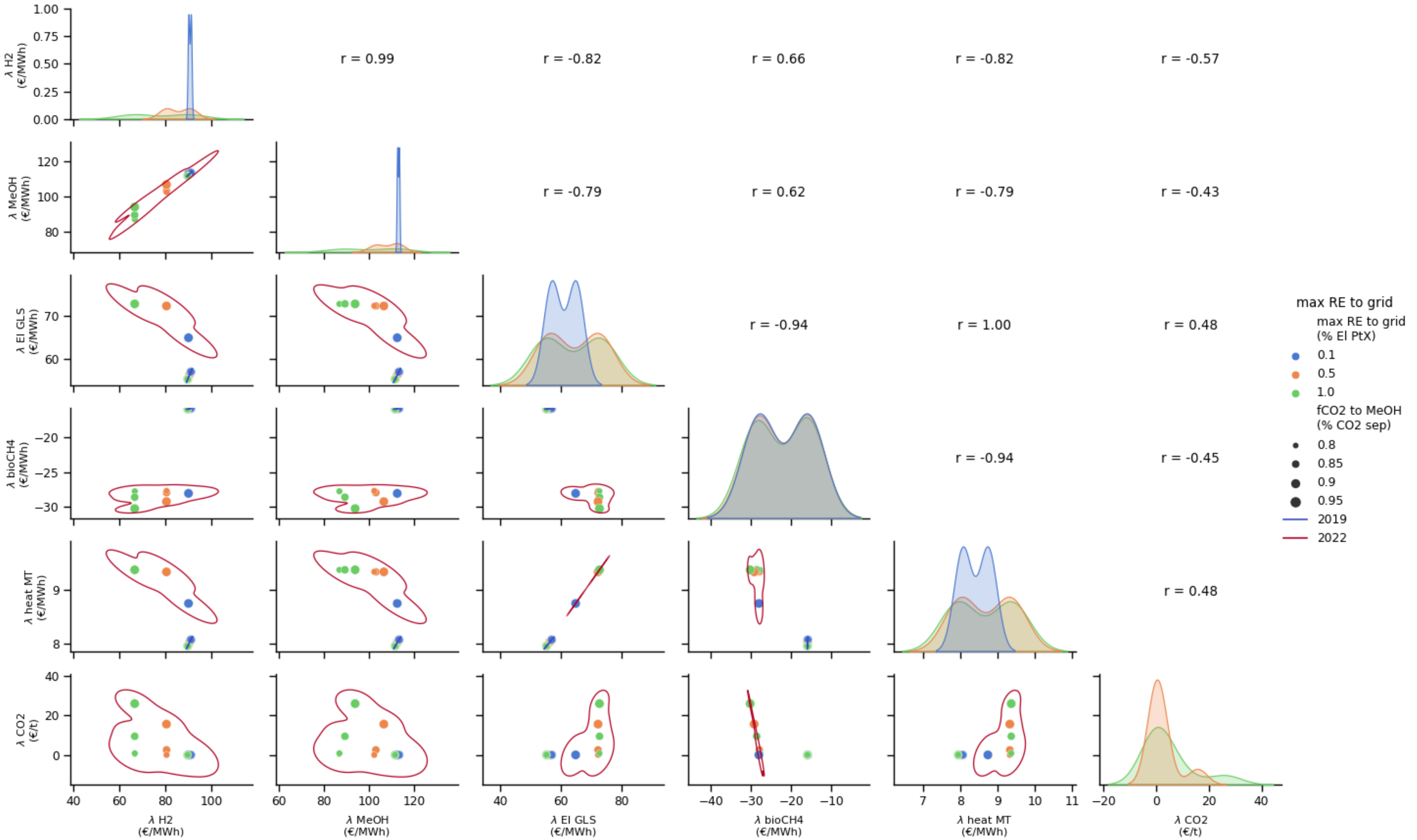

**Figure D3** – *H₂ to grid* scenario: correlation between shadow prices in the optimal design of the PtX hub. $CO_2$ tax: 150 (€/t), DH and biochar production not enabled.

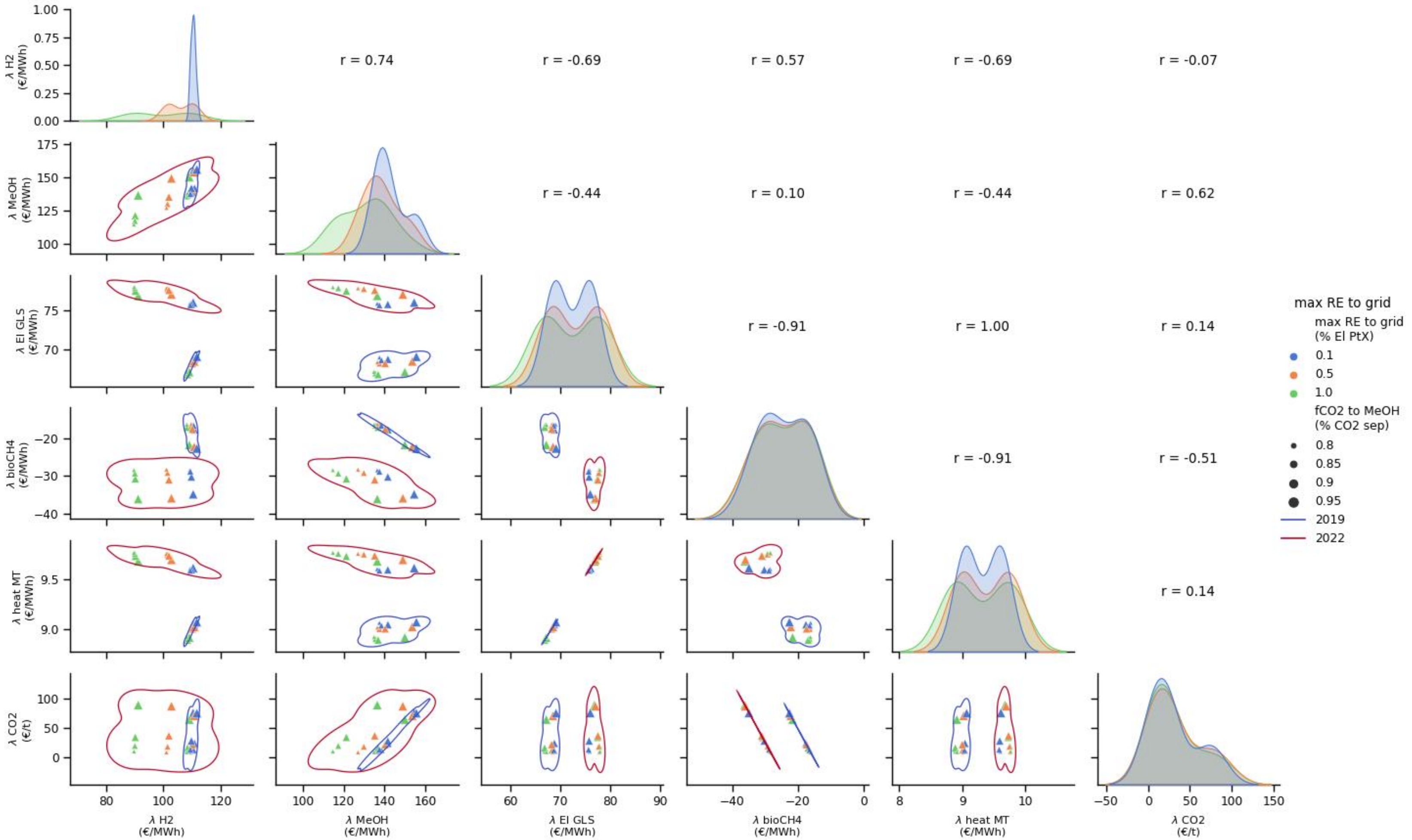

**Figure D4** – *MeOH standalone* scenario: correlation between shadow prices in the optimal design of the PtX hub. $CO_2$ tax: 150 (€/t), DH and biochar production not enabled.

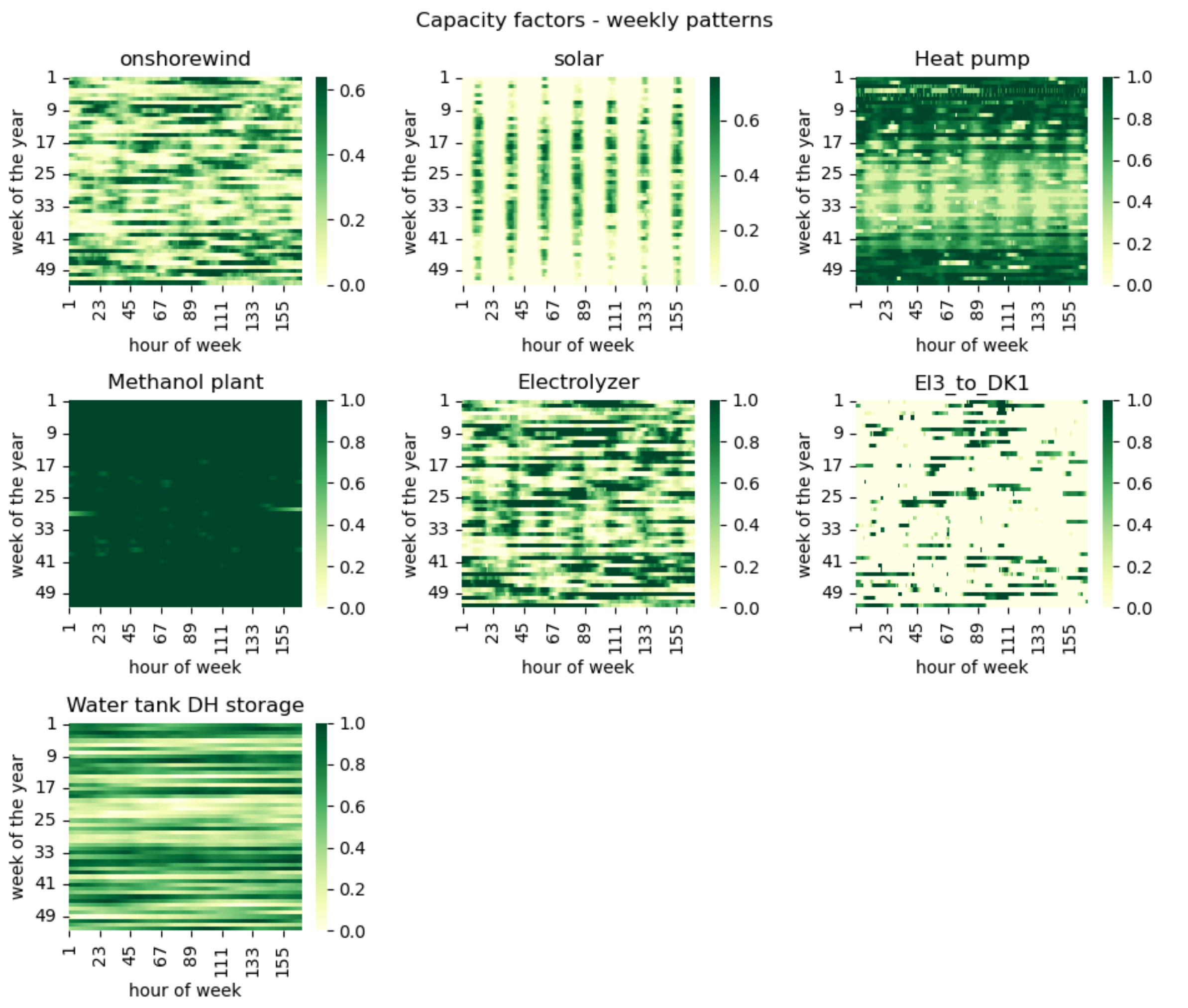

**Figure D5** – *$H_2$ to Grid* case: example of optimal operation in the PtX hub. Rate of $CO_2$ to methanol: 0.9, $CO_2$ tax: 150 (€/t), DH enabled, biochar not enabled, energy prices 2019.

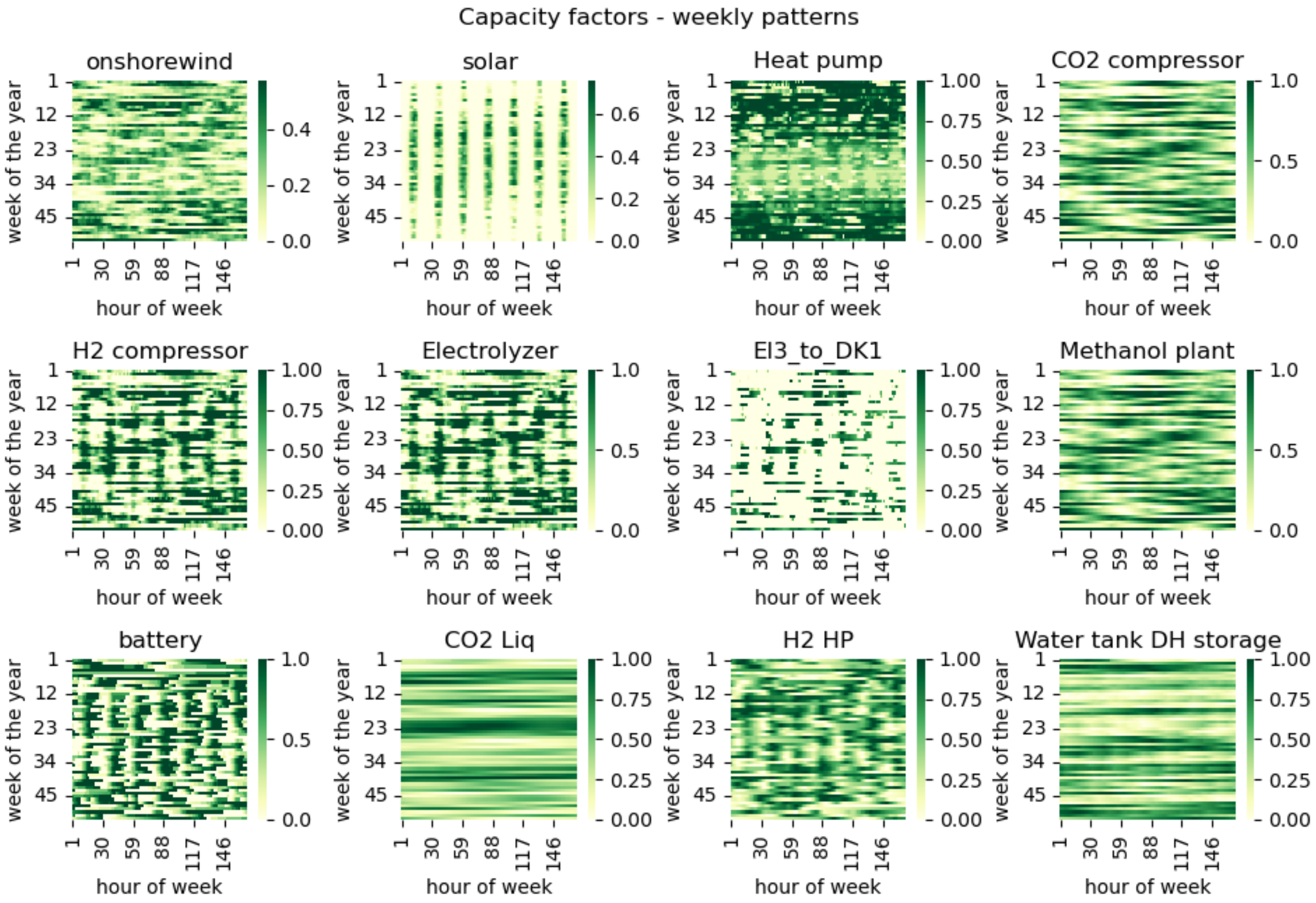

**Figure D6** – *MeOH standalone* case: example of optimal operation in the PtX hub. Rate of $CO_2$ to methanol: 0.9, $CO_2$ tax: 150 (€/t), DH enabled, biochar not enabled, energy prices 2019.

## Appendix E - Superstructure of the optimization model

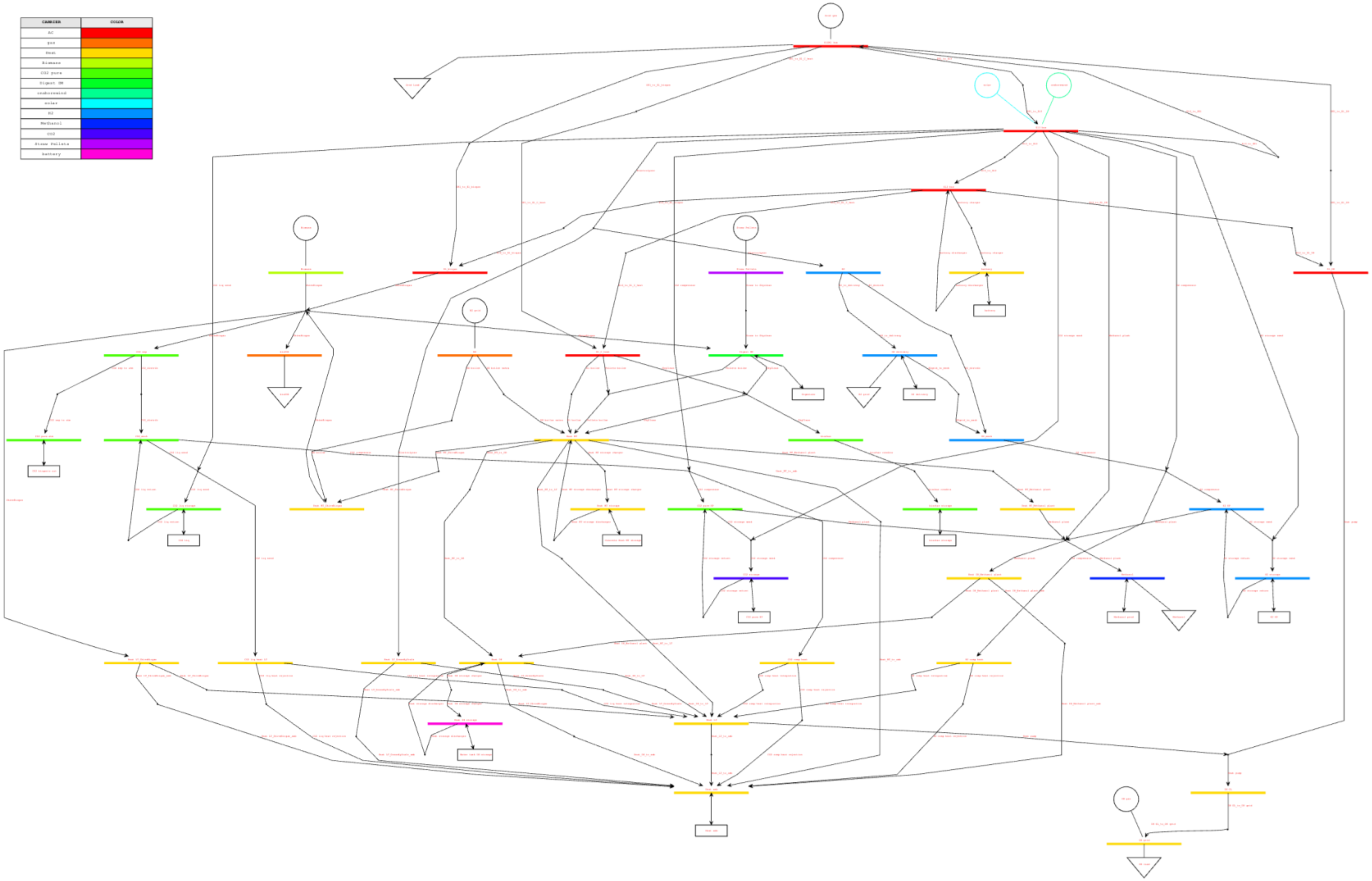

**Figure E1 –** Superstructure of the PyPSA optimization model including all technology alternatives before optimization.

## Appendix F - Legislative Framework in EU for RFNBOs

### Renewable hydrogen production

The RED outlines criteria for classifying hydrogen generated from electricity sourced from renewable energy installations (e.g. wind or solar parks) through a power purchase agreement (PPA) as 'green'. This classification is contingent on three cumulative principles: additionality, temporal correlation, and geographic correlation. A transitional phase is allowed for the additionality and temporal correlation (Fig. 2).

**Additionality**: This stipulates that the renewable energy production facility must not have commenced operations earlier than 36 months prior to the establishment of the hydrogen plant. This ensures that the 'green' hydrogen is derived from newly established, 'additional' capacity, rather than from pre-existing renewable energy sources. This requirement is based on the rationale that using renewable energy from existing installations (which might have otherwise supplied renewable energy to other customers) would lead to increased electricity production from fossil fuels (such as coal or gas-fired plants) to fulfill the demand of those customers. Furthermore, it is generally preferred that the renewable energy production installation has not received operational or investment subsidies. However, the additionality principle does apply during the transitional phase i.e. until 2038 for plants coming in operation before 2028.

**Temporal Correlation**: The requirement for temporal correlation specifies that renewable hydrogen must be produced in one of the following scenarios:

1. During the same one-hour period as the renewable electricity generated from the renewables (PPA).
2. From renewable electricity sourced from a newly established storage asset if this asset is situated behind the same network connection point as the electrolyser or the renewable electricity generation facility, and it has been charged during the same one-hour period in which the renewable electricity was produced.
3. During a one-hour period in which the price of electricity, is either lower than or equal to EUR 20 per MWh, or lower than 0.36 times the price of an allowance to emit one ton of carbon dioxide equivalent.

While the hourly synchronization of production and offtake is intended to prevent the use of fossil-based electricity it has faced criticism from the industry and lead, the EC to introduce a transition regime that will be in effect until 1 January 2030. During this period, temporal matching will occur monthly rather than an hourly one.

**Geographic Correlation**: This requirement dictates that the renewable energy production facility, and the hydrogen plant should ideally be situated in the same bidding zone, or in the offshore bidding zone that is linked to a bidding zone.

### Carbon tax

Explicit carbon taxes have been increasing in EU for all sectors and are subject to a continuous discussion and update for the legislation. Denmark participates in the EU emissions trading system (ETS) (OECD, 2018[1]) for power generation and manufacturing industries where carbon credits are traded at about 90€/t$CO_2$ as (September 2023). Facilities that are covered by the ETS do not to pay the explicit carbon tax, however plants for district heating are subject to the CO2 tax, irrespective of whether they are also covered by the EU ETS. Denmark as recently planned and to increase its carbon tax up to 750 (DKK/t$CO_2$) for companies not part of ETS scheme or additional 350 (DKK/t$CO_2$) for companies included in the ETS scheme by 2030, as part of the plan for expanding renewable power in the country.

For maritime industry two European Union (EU) carbon-pricing initiatives have been released serves as a notable example. From 2024, emissions from ships (exceeding 5000 gross tons) will be disclosed through the EU's Monitoring, Reporting, and Verification (MRV) system. These emissions will also fall under the scope of ETS system hence the vessels included will be obligated to procure allowances and to offset their GHG emissions for intra-EEA voyages and while stationed at EEA ports. Starting in 2025, 40% of the $CO_2$ emissions will be subject to the ETS, with a progressive increase to 100% by 2027 (with inclusion of $CH_4$ and $N_2O$ emissions). A second initiative, FuelEU Maritime, part of REPower EU, will be gradually enforced from 2025 onwards. This regulation establishes targets for diminishing the annual average GHG intensity of the energy utilized by a collective fleet or pool of ships. The reduction is initial modest, at -2% in 2025 (relative to a 2020 baseline), progressing to -6% in 2030 and -14.5% in 2035, ultimately reaching -80% by 2050.

# Nomenclature

| | Description | Unit |
|---|---|---|
| | **Indices** | |
| $t$ | Dispatching periods | |
| $s$ | Generators | |
| $n$ | Buses | |
| $ld$ | Loads | |
| $l$ | Links | |
| $st$ | Stores | |
| $el$ | electricity | |
| $NG$ | Natural gas | |
| $DH$ | District heating | |
| | **Superscripts** | |
| $ref$ | reference year | |
| $SC$ | scenario | |
| $in$ | purchase | |
| $out$ | sales | |
| | | |
| | **Parameters** | |
| $\overline{g}_{\bar{l}}$ | Ramp-up power constraint for dispatchable technologies (s) | MW/t or (t/h)/t |
| $\overline{g}_{\underline{l}}$ | Ramp-down power constraint for dispatchable technologies (s) | MW/t or (t/h)/t |
| | | |
| | **Variables** | |
| $G_s$ | Power capacity of technology *s* | MW or t/h |
| $c_s$ | Fix annualized costs for power capacity of technology *s* | €/MW or €/(t/h) |
| $E_s$ | Store Energy capacity of technology *s* | MWh or t |
| $\hat{c}_{st}$ | Fix annualized costs for power capacity of storage technology *s* | €/ MWh or €/t |
| $g_{s,t}$ | Power dispatch of technology *s* at time *t* | MW or t/h |
| $o_{s,t}$ | Variable cost of technology *s* at time *t* | €/MW or €/(t/h) |
| $d_{ld,t}$ | Demand of load *d* at time *t* | MW or t/h |
| $\alpha_{lk,t}$ | Efficiency and flow direction on the bus for link *lk* at time t | ± (MW/MW) or ± (t/h / t/h) |
| $f_{lk,t}$ | Energy or material flow for link *lk* at time t | MW or t/h |
| $\underline{f}_{lk,t}$ | Minimum power availability of link | ± (MW/MW) or ± (t/h / t/h) |
| $\overline{f}_{lk,t}$ | Maximum power availability of link | ± (MW/MW) or ± (t/h / t/h) |
| $\underline{g}_{gn,t}$ | Minimum power availability of generator | ± (MW/MW) or ± (t/h / t/h) |
| $\overline{g}_{gn,t}$ | Maximum power availability of generator | ± (MW/MW) or ± (t/h / t/h) |
| $e_{st,t}$ | Energy level in a storage *s* at time *t* | MWh or t |
| $\overline{g}_{\bar{l}}$ | Ramp-up power constraint for dispatchable technologies | MW/t or (t/h)/t |
| $\overline{g}_{\underline{l}}$ | Ramp-down power constraint for dispatchable technologies | MW/t or (t/h)/t |
| $\lambda_{n,t}$ | KKT multiplier (shadow price) for energy balance at bus *n* at time *t* | €/MW or €/(t/h) |
| $\mu$ | KKT multiplier (shadow price) for inequality constraint for technology s or st | €/MWh or €/t |
| $p_{el,p,t}$ | Total electricity price for purchase from the grid | €/MWh (time serie) |
| $p_{elREF,t}$ | Spot price for electricity during the reference year | €/MWh (time serie) |
| $TF_{el,p}$ | Total tariff for purchase of electricity | €/MWh |
| $em_t$ | Emission intensity of electricity production in | $tCO_2$/MWh (time serie) |
| $p_{CO2}$ | $CO_2$ tax (dependent on scenario) | |